\documentclass[aps,prd,onecolumn,groupedaddress,preprintnumbers,superscriptaddress,showpacs,nofootinbib]{revtex4}
\usepackage{amsmath,amssymb,subfigure}
\usepackage[dvips]{graphicx}

\usepackage{color}

\newcommand{\6}{\partial}

\newcommand{\K}{\mathcal{K}}

\begin{document}

\title{%
Gravitational wave signal from massive gravity
}

\author{A.~Emir~G\"umr\"uk\c{c}\"uo\u{g}lu} \email{emir.gumrukcuoglu@ipmu.jp}
\affiliation{Kavli Institute for the Physics and Mathematics of the Universe, Todai Institutes for Advanced Study, University of Tokyo, 5-1-5 Kashiwanoha, Kashiwa, Chiba 277-8583, Japan}
\author{Sachiko~Kuroyanagi} \email{skuro@resceu.s.u-tokyo.ac.jp}
\affiliation{Research Center for the Early Universe (RESCEU), Graduate School of Science,
The University of Tokyo, Tokyo 113-0033, Japan}
\author{Chunshan~Lin} \email{chunshan.lin@ipmu.jp}
\affiliation{Kavli Institute for the Physics and Mathematics of the Universe, Todai Institutes for Advanced Study, University of Tokyo, 5-1-5 Kashiwanoha, Kashiwa, Chiba 277-8583, Japan}
\author{Shinji~Mukohyama} \email{shinji.mukohyama@ipmu.jp} 
\affiliation{Kavli Institute for the Physics and Mathematics of the Universe, Todai Institutes for Advanced Study, University of Tokyo, 5-1-5 Kashiwanoha, Kashiwa, Chiba 277-8583, Japan}
\author{Norihiro~Tanahashi} \email{tanahashi@ms.physics.ucdavis.edu}
\affiliation{Department of Physics, University of California, Davis, CA 95616, USA}

\pacs{04.50.Kd, 04.30.-w}

\date{\today}
\preprint{IPMU12-0164}
\begin{abstract}
 We discuss the detectability of gravitational waves with a time
 dependent mass contribution, by means of the stochastic gravitational
 wave observations. Such a mass term typically arises in the
 cosmological solutions of massive gravity theories. We conduct the
 analysis based on a general quadratic action, and thus the results 
 apply universally to any massive gravity theories in which modification
 of general relativity appears primarily in the tensor modes. The
 primary manifestation of the modification in the gravitational wave spectrum
 is a sharp peak. The position and height of the peak carry information
 on the present value of the mass term, as well as the duration of the
 inflationary stage. We also discuss the detectability of such a
 gravitational wave signal using the future-planned gravitational wave
 observatories. 
\end{abstract}

\maketitle

\section{Introduction}

Since the pioneering model of massive gravity was proposed by Fierz and
Pauli \cite{Fierz:1939ix}, numerous attempts have been made to introduce
a non-zero mass to graviton. This issue has been attracting a great deal
of interest, partly because graviton mass may provide an alternative
explanation for the acceleration of our universe. Namely, instead of
attributing it to the existence of dark energy whose origin is still
unknown, the acceleration may be simply due to modification of gravity.

Until recently, however, it has been thought that such modification by
non-zero graviton mass is extremely difficult, typically leading to the 
emergence of ghost degrees of freedom and various pathologies. As first
shown by Boulware and Deser~\cite{Boulware:1973my}, the source of these
difficulties is associated with a helicity-$0$ mode in the gravity
sector, which is absent at the quadratic order of the Fierz-Pauli
massive gravity but revives at the higher order and acts as a
ghost, dubbed the {\it Boulware-Deser (BD) ghost}. 
A remedy for this difficulty was recently proposed by
Refs.~\cite{deRham:2010ik,deRham:2010kj}, in which the would-be BD ghost
is eliminated by adding higher-order correction terms order by order and by
resumming them into a three parameter expression. Non-existence of the BD ghost
in this theory was discussed and proved in 
Refs.~\cite{Hassan:2011hr,deRham:2011rn,deRham:2011qq,Hassan:2011tf,Hassan:2011ea,Mirbabayi:2011aa,Golovnev:2011aa, Kluson}.
Various solutions in the theory were studied in  
Refs.~\cite{deRham:2010tw, Koyama:2011xz, Nieuwenhuizen:2011sq, Koyama:2011yg, Chamseddine:2011bu, D'Amico:2011jj, Gumrukcuoglu:2011ew, Koyama:2011wx, Comelli:2011wq, Volkov:2011an, vonStrauss:2011mq, Comelli:2011zm, Berezhiani:2011xx,Brihaye:2011aa,Kobayashi:2012fz}.
Therefore this nonlinear theory of massive gravity can be considered as
a nonlinear completion of the Fierz-Pauli theory, which is the simplest
and the oldest linear massive gravity theory. Having this in
mind, we consider it quite natural to suppose that other massive gravity
theories or, more generally, Higgs phases of gravity might find their
nonlinear completions in the future.

Such examples include, but are not restricted to, ghost
condensate~\cite{ArkaniHamed:2003uy,ArkaniHamed:2005gu} and 
Lorentz-violating massive gravity
theories~\cite{Rubakov:2004eb,Dubovsky:2004sg}. In these examples, 
propagating degrees of freedom do not form a representation of
$4$-dimensional Poincar\'e symmetry even in the exact Minkowski
background. Instead, they form representations of $3$-dimensional 
rotational and translational symmetries. As a result, physical degrees of
freedom are classified into scalar, vector and tensor parts, according
to their transformation properties under the $3$-dimensional spatial 
rotation. Modes in different classes may behave rather differently. For
instance, in the case of ghost condensate in Minkowski background,
propagating degrees of freedom in the gravity sector are two tensor
modes and one scalar mode. The two tensor modes behave exactly like
those in general relativity (GR) while the scalar mode has an unusual
dispersion relation, i.e.\ non-relativistic dispersion relation without a
mass gap. On the other hand, in a particular class of models among those 
proposed in \cite{Dubovsky:2004sg}, propagating degrees of freedom in
the gravity sector are two tensor modes only, and their dispersion
relation is massive and relativistic. From the point of view of the
symmetry, i.e.\ $3$-dimensional spatial rotational symmetry, we expect
that there should be even wider classes of models in which scalar, 
vector, and tensor parts behave differently.

In the present paper, for definiteness we mainly focus on a class of
massive gravity models in which propagating degrees of freedom in
gravity sector are only two tensor modes with a massive dispersion
relation. In this case, the scalar and vector sectors behave exactly
like in GR and the only modification emerges in the tensor sector: the
dispersion relation of gravitational waves acquires an effective mass,
which in general can be time-dependent. We shall discuss an attempt
towards a nonlinear completion of such massive gravity models, but in
most part of the present paper we shall consider this class of models as
a purely phenomenological one to be constrained or probed by
observations and experiments. We conduct the analysis based on a general
quadratic action for tensor modes, and thus our results apply not only
to the theories with precisely two propagating gravitational modes but
also to a more general class of theories in which modification of
gravity appears primarily in the tensor modes.

Being different from GR only in the tensor sector, observation of
gravitational waves and/or their imprints would be the most efficient
probe for the model. Therefore, in this paper we address the
detectability of the effective mass of gravitational waves by means of
the stochastic gravitational wave observations.

The rest of this paper is organized as follows. After describing our
model in Sec.~\ref{Sec:basics}, we discuss an attempt towards nonlinear
completion of the model in Sec.~\ref{Sec:nonlinear}. This consideration
at the very least shows that the structure of the model considered in
the present paper is not forbidden by the symmetry. However, readers who
are interested only in purely phenomenological aspects can safely skip
Sec.~\ref{Sec:nonlinear}. We then derive analytic formulas for the
power-spectrum of stochastic gravitational waves and discuss how to read
off information about the effective mass of gravitational waves in
Sec.~\ref{Sec:power}.  In Sec.~\ref{Sec:example} we show some numerical
results and make comparison with the analytical results of the previous
section. Sec.~\ref{Sec:Summary} is devoted to a summary of the results
and discussions.

\section{Model description}
\label{Sec:basics}

In the present paper we shall consider linear massive gravity models
that respect the symmetry of Friedmann-Robertson-Walker (FRW) universe, 
i.e.\ the spatial homogeneity and isotropy. Metric perturbations are thus
decomposed into scalar, vector and tensor parts, according to the
transformation properties under the $3$-dimensional spatial rotation.

We restrict our considerations to a class of models in which
the background solution is given by the ordinary FRW universe (with or without 
self-acceleration) and 
only the two tensor modes in the gravity sector are propagating on this background.
For tensor-type perturbations on the FRW background, the
metric can be written as 
\begin{equation}
 g_{\mu\nu} dx^\mu dx^\nu
= -N(t)^2 dt^2 + a(t)^2 
  \left[\Omega_{ij}(x^k) + \gamma_{ij}\right]dx^idx^j,
  \quad
\Omega_{ij}(x^k) = \delta_{ij}
+ \frac{K\delta_{il}\delta_{jm}x^lx^m}
{1-K\delta_{lm}x^lx^m}. 
\end{equation}
where $\gamma_{ij}$ is the tensor-type perturbations satisfying
\begin{equation}
 D^i\gamma_{ij} = 0, \quad \Omega^{ij}\gamma_{ij} = 0. 
\end{equation}
Here, $\Omega^{ij}=(\Omega^{-1})^{ij}$, $D^i=\Omega^{ij}D_j$ and 
$D_j$ is the spatial covariant derivative compatible with the metric
$\Omega_{ij}$.

By assumption, the quadratic action for the perturbation should be
invariant under the spatial rotation and translation. Hence we can use
the following ingredients to construct the quadratic action describing
the dynamics of the tensor perturbation:  
\begin{equation}
 Ndt, \quad \sqrt{\Omega}d^3\vec{x}, \quad \Omega_{ij}, \quad D_i, 
  \quad \gamma_{ij}, \quad \frac{1}{N}\dot{\gamma}_{ij}, 
\end{equation}
provided that spatial indices are properly contracted to form scalar
quantities. Here, $\Omega\equiv\det\Omega_{ij}$ and an overdot
represents derivative with respect to the time coordinate $t$. Since the
time coordinate $t$ is invariant under the spatial rotation and
translation, we can also use general functions of $t$ and their derivatives.

We also demand that the action should include only up to second time 
derivatives of $\gamma_{ij}$ in order to avoid ghosts in the linear
level and that the system respects the spatial parity invariance.

Thus, after integration by parts, one can show that the general
quadratic action is of the form
\begin{equation}
 I^{(2)}_{tensor} = \frac{M_{Pl}^2}{8}\int dt d^3 x
\, Na^3\sqrt{\Omega}\, F(t)
\left[\frac{1}{N^2}\dot\gamma^{ij}\dot\gamma_{ij}
 +\gamma^{ij}
 \left(\sum_{n=0}^{\infty}c_n(t)\frac{\triangle^n}{a^{2n}}\right)
 \gamma_{ij} \right],
\label{action}
\end{equation}
where $\gamma^{ij}\equiv\Omega^{ik}\Omega^{jl}\gamma_{kl}$, 
$\triangle\equiv D^iD_i$ is the Laplacian operator associated with
$\Omega_{ij}$, $F(t)$ and $c_n(t)$ ($n=0,1,\cdots$) are general
functions of the time coordinate $t$. Note that $F(t)$ must be positive
definite in order to avoid appearance of ghost and strong coupling. We
can then set $F(t)=1$ by a field redefinition, to be more precise, by a
conformal transformation. Finally, in order to describe low energy
phenomena such as cosmological evolution of gravitational waves at late
time, we truncate the series expansion (the sum over $n$) at the second
order spatial derivatives and obtain
\begin{equation}
 I^{(2)}_{tensor} = \frac{M_{Pl}^2}{8}\int dt d^3 x
\, Na^3\sqrt{\Omega}
\left[\frac{1}{N^2}\dot\gamma^{ij}\dot\gamma_{ij}
+ \frac{c_g^2(t)}{a^2}\gamma^{ij}\left(\triangle-2K\right)\gamma_{ij}
- M_{GW}^2(t)\gamma^{ij}\gamma_{ij}
 \right], \label{I2}
\end{equation}
where we have set $F=1$ by a field redefinition (or a conformal
transformation) and we have defined 
\begin{equation} 
 c_g^2(t)\equiv c_1(t), \quad 
  M_{GW}^2(t)\equiv  -c_0(t) - \frac{2Kc_1(t)}{a^2}.
\label{Mgwdef_general}
\end{equation}
Note that physical meaning of $c_g$ and $M_{GW}$ are the sound speed and
effective mass of gravitational waves.

Taking advantages of the background symmetry, it is convenient to expand 
$\gamma_{ij}$ as
\begin{equation}
 \gamma_{ij}
= \sum_{\lambda}
\int k^2 dk \; \gamma_{k,\lambda}
Y_{ij}^\lambda
\bigl(\vec k, \vec x\bigr),
\end{equation}
where $k^2\equiv \Omega_{ij}\vec k^i \vec k^j$, 
$\lambda$ denotes the helicity state and
$Y_{ij}^\lambda$ is the tensor harmonics satisfying
\begin{equation}
\left(\triangle+ k^2\right)Y_{ij}^\lambda =0,
\quad
D^iY_{ij}^\lambda=0,
\quad
\Omega^{ij} Y_{ij}^\lambda = 0,
\quad
\int d^3 x\sqrt\Omega \; \Omega^{ij} \Omega^{kl}
Y_{ik}^\lambda \bigl(\vec k, \vec x\bigr)
Y_{jl}^{\lambda'} \bigl(\vec k', \vec x\bigr)
=
32\pi^2 \delta_{\lambda\lambda'}
\delta^3\bigl(\vec k + \vec k'\bigr).
\end{equation}
The equation of motion for $\gamma_{k,\lambda}$ is then 
\begin{equation}
\bar \gamma_k''+\left(
c_g^2k^2 + a^2 M_{GW}^2 - \frac{a''}{a} + 2Kc_g^2
\right)\bar \gamma_k = 0, \quad
\bar \gamma_k \equiv a\gamma_k,
\label{eqtensor}
\end{equation}
where the prime (${}'$)
denotes derivative with respect to the conformal time 
$\eta$ defined by $d\eta \equiv N dt /a$. 
Since the equation of motion is identical for both polarizations, we omit the index $\lambda$ hereafter.

\section{Attempts toward nonlinear completion}
\label{Sec:nonlinear}

The nonlinear theory of massive gravity recently proposed by
Refs.~\cite{deRham:2010ik,deRham:2010kj} eliminates the renowned 
BD ghost by construction and thus can be considered as a nonlinear
completion of the Fierz-Pauli theory, which is the simplest and the
oldest among all linear massive gravity theories. Having this in mind,
we consider it quite possible that the phenomenological model described
in the previous section might also find its nonlinear completions in the
future.

In this section we discuss an attempt towards such a nonlinear
completion. In particular, we shall derive the model with $c_g=1$
precisely, based on a non-trivial background in the nonlinear theory of
massive gravity. Unfortunately, this construction is purely classical
and fails at the quantum level~\cite{DGM}.\footnote{Adopting a different 
approach, Ref.~\cite{D'Amico:2012pi} achieved a similar conclusion.} Nonetheless, at the very
least it shows that the structure of the model considered in the present
paper, with $c_g=1$, is not forbidden by symmetry. Readers who are
interested only in purely phenomenological aspects can safely skip this
section.

The covariant action of the nonlinear massive gravity is constructed out
of a $4$-dimensional metric $g_{\mu\nu}$ and four scalar fields
$\varphi^a$ ($a=0,1,2,3$) called {\it St\"uckelberg fields}. The
St\"uckelberg fields enter the action only through the tensor
$f_{\mu\nu}$ defined as 
\begin{equation}
 f_{\mu\nu} = \bar{f}_{ab}(\varphi^c)
  \partial_{\mu}\varphi^a\partial_{\nu}\varphi^b,
\end{equation}
where $\bar{f}_{ab}(\varphi^c)$ is a non-degenerate, second-rank
symmetric tensor in the field space. The spacetime metric $g_{\mu\nu}$
and the tensor $f_{\mu\nu}$ are often called {\it physical metric} and
{\it fiducial metric}, respectively.

The gravity action is the sum of the Einstein-Hilbert action $I_{EH,\Lambda}$ (including a bare
cosmological constant $\Lambda$) and the graviton mass term $I_{mass}$ specified
below. Adding the matter action $I_{matter}$, the total action is 
\begin{equation}
 I = I_{EH,\Lambda}[g_{\mu\nu}] + I_{mass}[g_{\mu\nu},f_{\mu\nu}] 
  + I_{matter}[g_{\mu\nu},\sigma_I], 
\end{equation}
where 
\begin{eqnarray}
 I_{EH,\Lambda}[g_{\mu\nu}] & = & 
  \frac{M_{Pl}^2}{2}\int d^4x\sqrt{-g}(R-2\Lambda),
  \\
 I_{mass}[g_{\mu\nu},f_{\mu\nu}] & = & 
  M_{Pl}^2m_g^2\int d^4x\sqrt{-g}\,
  ( {\cal L}_2+\alpha_3{\cal L}_3+\alpha_4{\cal L}_4), 
  \label{eqn:Imass}
\end{eqnarray}
and $\{\sigma_I\}$ ($I=1,2,\cdots$) represent generic matter fields. Each
contribution in the mass term $I_{mass}$ is constructed as 
\begin{eqnarray}
 {\cal L}_2 & = & \frac{1}{2}
  \left(\left[{\cal K}\right]^2-\left[{\cal K}^2\right]\right)\,, \nonumber\\
 {\cal L}_3 & = & \frac{1}{6}
  \left(\left[{\cal K}\right]^3-3\left[{\cal K}\right]\left[{\cal K}^2\right]+2\left[{\cal K}^3\right]\right), 
  \nonumber\\
 {\cal L}_4 & = & \frac{1}{24}
  \left(\left[{\cal K}\right]^4-6\left[{\cal K}\right]^2\left[{\cal K}^2\right]+3\left[{\cal K}^2\right]^2
   +8\left[{\cal K}\right]\left[{\cal K}^3\right]-6\left[{\cal K}^4\right]\right)\,,
\label{lag234}
\end{eqnarray}
where the square brackets denote trace operation and 
\begin{equation}
{\cal K}^\mu _\nu = \delta^\mu _\nu 
 - \left(\sqrt{g^{-1}f}\right)^{\mu}_{\ \nu}\,.
\label{Kdef}
\end{equation}
The square-root in this expression is the positive definite matrix
defined through 
\begin{equation}
 \left(\sqrt{g^{-1}f}\right)^{\mu}_{\ \rho}
  \left(\sqrt{g^{-1}f}\right)^{\rho}_{\ \nu}
  = f^{\mu}_{\ \nu}\ (\equiv g^{\mu\rho}f_{\rho\nu}).
  \label{eqn:square-root}
\end{equation}

Note that the fiducial metric is not dynamical but fixed by the
theory. In other words, the form of the fiducial metric is a part of
definition of the theory. Of course, an apparent form of the fiducial
metric is subject to changes under redefinition and dynamics of
the St\"uckelberg fields. However, such changes are equivalent to coordinate
transformations of the metric in the field space and thus do not change
geometrical properties such as the curvature tensor in the field
space. For example, if we demand the maximal symmetry, i.e.\ Poincar\'e, de
Sitter or anti-de Sitter symmetry, in the field space then the fiducial
metric must be Minkowski, de Sitter or anti-de Sitter metric,
respectively, regardless of appearances. 

In the present paper we consider a general
fiducial metric of the FRW type, specified by the lapse function 
$n(\varphi^0)$, the scale factor $\alpha(\varphi^0)$ and the spacial
curvature constant $K$: 
\begin{equation}
f_{\mu\nu} =
- n^2(\varphi^0)\6_\mu\varphi^0 \6_\nu\varphi^0
+ \alpha^2(\varphi^0)\Omega_{ij}(\varphi^k)
\6_\mu\varphi^i \6_\nu\varphi^j,
\qquad
\Omega_{ij}(\varphi^k) = \delta_{ij}
+ \frac{K\delta_{il}\delta_{jm}\varphi^l\varphi^m}{
1-K\delta_{lm}\varphi^l\varphi^m}.
\label{FRWfiducial}
\end{equation}
Needless to say, this class of fiducial metrics include Minkowski, de
Sitter and anti-de Sitter fiducial metrics as special cases. For
example, a Minkowski fiducial metric can be rewritten as an open FRW
form and allows non-trivial cosmology~\cite{Gumrukcuoglu:2011ew}. Note,
however, that the Minkowski fiducial metric does not allow flat nor
closed FRW cosmology~\cite{D'Amico:2011jj}. Similarly, an anti-de Sitter 
fiducial metric allows non-trivial open FRW cosmology but does not allow 
flat nor closed ones (see Appendix~\ref{App:fiducials}). On the other
hand, non-trivial open, flat, closed FRW cosmologies can be realized
with a de Sitter fiducial metric (see Appendix~\ref{App:fiducials}, and
see also Ref.~\cite{Gumrukcuoglu:2011zh,Langlois:2012hk} for related
discussions).

In this section we consider a general FRW background plus
perturbations. 
\begin{equation}
 g_{\mu\nu} = g^{(0)}_{\mu\nu} + \delta g_{\mu\nu},
  \qquad
  \varphi^a = x^a + \pi^a + \frac{1}{2}\pi^b\partial_b\pi^a +
  O(\epsilon^3), \label{eqn:def-perturbation}
\end{equation}
where 
\begin{equation}
 g^{(0)}_{\mu\nu} dx^\mu dx^\nu = -N(t)^2 dt^2 + a(t)^2 
  \Omega_{ij}(x^k)dx^idx^j,
\qquad
\Omega_{ij}(x^k) = \delta_{ij}
+ \frac{K\delta_{il}\delta_{jm}x^lx^m}
{1-K\delta_{lm}x^lx^m}. 
\label{physical}
\end{equation}
Here, the perturbation $\pi^a$ of the St\"uckelberg fields was defined
through the exponential map, following~\cite{Gumrukcuoglu:2011zh}. The
background equations of motion lead to
\begin{equation}
3H^2 + \frac{3K} {a^2}
= \Lambda_\pm + \frac{1}{M_{Pl}^2} \rho
,
\qquad
-\frac{2\dot H}{N} + \frac{2K}{a^2} =
\frac{1}{M_{Pl}^2}\left(\rho+P\right),
\qquad
\frac{\alpha(t)}{a(t)} = X_\pm,
\label{Friedmann}
\end{equation}
where a dot represents derivative with respect to $t$,  
$H\equiv\dot{a}/(Na)$ is the Hubble expansion rate of the background
physical metric, $\rho$ and $P$ are the matter energy density and
pressure, and 
\begin{equation}
X_\pm \equiv \frac{
1 + 2\alpha_3 + \alpha_4 \pm \sqrt{1+\alpha_3+\alpha_3^2-\alpha_4}
}{\alpha_3+\alpha_4},
\qquad
\Lambda_\pm = -m_g^2 (1-X_\pm)[
3-X_\pm + \alpha_3 (1-X_\pm)
].
\label{effectiveLambda}
\end{equation}
(See Ref.~\cite{Gumrukcuoglu:2011zh} or Appendix~\ref{App:fiducials}
for derivation.) Note that the
background physical metric follows the Friedmann equation with an
effective cosmological constant $\Lambda_{\pm}$ and thus the universe
exhibits self-acceleration if $\Lambda_{\pm}>0$.

 Thanks to the exponential map used to define $\pi^a$ in
 (\ref{eqn:def-perturbation}), the action for cosmological perturbations
 around the self-accelerating FRW background manifestly respects the
 background symmetry. This allows us to decompose the perturbations into
 scalar, vector and tensor parts, according to transformation properties
 under spatial diffeomorphism. A priori, because of the inclusion of
 four St\"uckelberg fields, we might expect cosmological perturbations
 in the gravity sector to consist of $6$ degrees of freedom: $2$
 tensor, $2$ vector and $2$ scalar modes. Actually, since the theory is
 constructed in such a way that the would-be BD ghost is completely
 excised, one would instead expect that cosmological perturbations in the
 gravity sector should consist of $5$ degrees of freedom: $2$ tensor,
 $2$ vector and $1$ scalar modes. Surprisingly enough, the
 gauge-invariant analysis of the quadratic action in
 \cite{Gumrukcuoglu:2011zh} has shown that the scalar and vector degrees
 have vanishing kinetic terms. Assuming these degrees are non-dynamical,
 we may integrate them out, then find that in the scalar and vector
 sectors, gauge-invariant variables constructed from metric and matter
 perturbations have exactly the same quadratic action as in general
 relativity.

 Hence, for linear perturbations around the self-accelerating FRW
 background, difference from general relativity arises only in the
 tensor sector. Assuming that there is no tensor-type contribution from
 matter fields $\sigma_I$ for simplicity, the quadratic action for the
 tensor sector is given by (\ref{I2}) with 
\begin{equation}
 c_g = 1, \quad
  M_{GW}^2=\pm(r-1)m_g^2X_\pm \sqrt{
  1+\alpha_3+\alpha_3^2- \alpha_4
  },
  \quad
  r = \frac{H}{X_\pm H_f}. 
\label{mgwdef}
\end{equation}
Here, $H\equiv \dot{a}/(Na)$ and $H_f\equiv \dot{\alpha}/(n\alpha)$
are Hubble expansion rates of the background physical metric and the 
fiducial metric, respectively. Note that $r$ and thus $M_{GW}$ are
time-dependent functions determined by the ratio $H/H_f$.

We have derived the model described in the previous section with
$c_g=1$. As already stated, the derivation is purely classical and fails
at the quantum level~\cite{DGM}. For this reason we do not consider this
construction as a realistic one. Nonetheless, at the very least this shows
that the structure of the model is not forbidden by symmetry. It may be
worthwhile focusing on different models such as
Refs.~\cite{Dubovsky:2004sg,Blas:2009my,deRham:2012kf} and/or
investigating cosmological perturbations around new classes of 
cosmological backgrounds such as those found in
Ref.~\cite{Koyama:2011wx,Kobayashi:2012fz,Gumrukcuoglu:2012aa,Motohashi:2012jd}, towards a successful nonlinear completion.

For example, in one of Lorentz-violating massive gravity
theories~\cite{Dubovsky:2004sg}, the symmetry 
$\phi^i\to \phi^i+\xi^i(\phi^0)$ ensures the existence of three
instantaneous modes in a fully nonlinear level, where $\xi^i$ ($i=1,2,3$)
are arbitrary functions of $\phi^0$. Invoking the additional scaling
symmetry $\phi^0\to \lambda\phi^0$, 
$\phi^i\to\lambda^{-\gamma}\phi^i$, and setting the $3$-dimensional
fiducial metric to be flat, one can show that linear perturbations
around a flat FRW background include only three physical modes in the
gravity sector (except for the three instantaneous modes): two tensor
modes representing massive gravitons and one scalar mode stemmed from
$\phi^0$~\cite{Dubovsky:2005dw}. The scalar mode is analogous to the
ghost condensate~\cite{ArkaniHamed:2003uy,ArkaniHamed:2005gu} in the
sense that it is massless and has vanishing sound speed. Since the
modification of gravity due to ghost condensate is rather
mild~\footnote{For example, if we do not include higher derivative terms
mentioned below, then the scalar mode does not modify gravity at 
linearized level in Minkowski background and thus the number of
propagating degrees of freedom in this case is effectively two. If the
higher derivative terms are included, then the scalar mode modifies
gravity but modification is still rather mild.}, this setup may be
considered as a good candidate for nonlinear completion. However, in the
ghost condensate, because of the vanishing sound speed, the leading
gradient terms come from higher derivative terms and thus the inclusion 
of higher derivative terms is essential for the stability of the system
at (relatively) short distances. When the above mentioned scaling
symmetry is imposed, this implies that higher derivative terms must be
introduced not only for $\phi^0$ but also for $\phi^i$. It is important
that those higher derivative terms for $\phi^i$ can in principle
introduce tiny time kinetic terms but that they do not lead to strongly
coupled (and possibly ghosty) new degrees of freedom in the regime of
validity of the effective field theory, even in curved backgrounds away
from the decoupling limit. To see this, note that the existence of
instantaneous modes implies that such a time kinetic term, if exists,
should have spatial derivatives acted on it. Therefore, as far as those
time kinetic terms are suppressed by the cutoff scale of the theory, the 
corresponding would-be new degrees of freedom have frequencies above the
cutoff. For this reason, it is expected that higher derivative terms do
not introduce any dangerous new degrees of freedom. It is certainly
worthwhile investigating this in concrete setups. Detailed analysis of
these issues is beyond the scope of the present  paper and thus we leave
it for the future.

\section{Stochastic gravitational wave spectrum}
\label{Sec:power}

To clarify the influence of $M_{GW}$ on the observable quantities, we
seek approximate solutions to Eq.~(\ref{eqtensor}) for a generic 
$M_{GW}(t)$, by considering super-horizon modes to be frozen and
sub-horizon modes to follow the WKB-type evolution. We then derive
approximate analytic formulas for the energy and power spectra. For
simplicity we set 
\begin{equation}
 c_g=1,
\end{equation}
throughout this and the forthcoming sections except for
Sec.~\ref{Sec:SummaryOfEnhancementFactor}, where we give a brief comment
on the case of a general $c_g$.

\subsection{Horizon re-entry}
\label{subsec:horizon-reentry}

>From the action (\ref{I2}) and the equation of motion (\ref{eqtensor})
with $c_g=1$, one can easily read off the dispersion relation as
\begin{equation}
 \omega^2 = \frac{k^2}{a^2} + M_{GW}^2, 
  \label{eqn:dispersion-relation}
\end{equation}
where $\omega$ is the effective frequency with respect to the proper
time $\tau$ defined by $d\tau=Ndt=ad\eta$. Here, since the comoving
momentum $k$ appears only through the combination $\sqrt{k^2+2K}$, we
have renamed this combination as $k$ so that $K$ does not appear
explicitly. We shall still call $k$ (after this redefinition) comoving
momentum. 
To consider the case that the graviton mass becomes in effect in the evolution 
history and also to simplify the analysis,
we assume that the mass squared $M_{GW}^2$,
which is an arbitrary function given by Eq.~(\ref{Mgwdef_general}),
dominates the right
hand side of (\ref{eqn:dispersion-relation}) at late time and that it
does not lead to instability, i.e. 
\begin{equation}
 \partial_t(a^2M_{GW}^2)> 0, \quad M_{GW}^2>0. 
  \label{mgwgrows}
\end{equation}
(See Sec.~\ref{Sec:example} and Appendix~\ref{App:fiducials}
for a simple example in which these
conditions actually hold.)

The frequency $\omega$ given by the dispersion relation
(\ref{eqn:dispersion-relation}) is relevant and the tensor mode
oscillates with this frequency if and only if it is sufficiently high
compared with the cosmic expansion, i.e.\ $\omega^2\gg H^2$. On the
other hand, when $\omega^2\ll H^2$, the equation of motion
(\ref{eqtensor}) allows a growing solution $\bar{\gamma}_k\propto a$ and
thus $\gamma_k\simeq const$. At a given moment of time, a mode with
$\omega^2\gg H^2$ is called {\it sub-horizon} and a mode with
$\omega^2\ll H^2$ is called {\it super-horizon}, as usual. For a given
mode with comoving momentum $k$, the time when 
$\omega^2\simeq H^2$ is called {\it horizon crossing}, again, as usual.

We suppose that scale-invariant super-horizon gravitational waves are
generated in the early universe, e.g.\ through inflation. In the present
paper we concentrate on the subsequent evolution of gravitational
waves, assuming for simplicity that the matter energy density and
pressure in the background equation of motion (\ref{Friedmann}) satisfy
the strong energy condition ($\rho+3P>0$). With this assumption,
$a^2\rho$ decreases as the universe expands. Since the curvature term
($3K/a^2$) is small (less than a few percent of $3H^2$) at present time,
this implies that the curvature term is negligible throughout the
evolution. Hence, before the effective cosmological constant
$\Lambda_{\pm}$ starts dominating the
universe,\footnote{\label{footnote:Lambda-dominance}We suppose that the
$\Lambda_{\pm}$ domination starts at around $z=O(1)$ and that the
evolution after that does not change gravitational wave spectra
significantly.} $a^2H^2$ decreases as the universe expands and, under
the assumption (\ref{mgwgrows}), super-horizon modes generated in the
early universe eventually cross the horizon and become sub-horizon. This
type of horizon crossing is called {\it horizon re-entry}, as usual.

For a mode with comoving momentum $k$, the time of horizon re-entry is
denoted as $t=t_k$ and defined as the solution to 
\begin{equation}
 \omega^2(t_k) = H^2(t_k).
\end{equation}
We denote the values of $a$ and $H$ at $t=t_k$ as $a_k$ and $H_k$,
respectively. For comparison, we define the time of horizon re-entry in
the massless case (GR) $t_k^{GR}$ by 
\begin{equation}
 \frac{k^2}{a^2} = H^2 \quad \mbox{at} \quad t=t_k^{GR},
\end{equation}
and denote the values of $a$ and $H$ at $t=t_k^{GR}$ as $a_k^{GR}$ and
$H_k^{GR}$, respectively.

\subsection{Classification of modes}
\label{subsec:classification}

As defined above, the modes re-enter the horizon when the effective
frequency $\omega$ given by (\ref{eqn:dispersion-relation}) exceeds the
Hubble expansion rate $H$. Thus, the discussion of the evolution of the
modes can be reduced to the determination of the time when either the mass
contribution $a\,M_{GW}$ or the comoving momentum $k$ exceeds the
comoving horizon scale $a\,H$. Hence, in this subsection, we discuss
the evolution of $a\,M_{GW}$ and $a\,H$ relative to $k$.

\begin{figure}[ht]
\includegraphics[width=.47\textwidth]{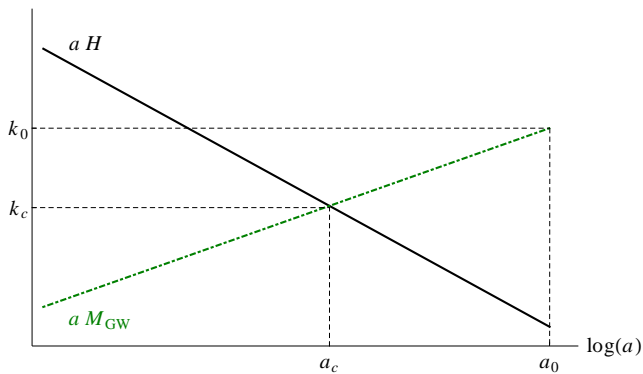}
\caption{Schematic plot for the evolution of $a\,M_{GW}$ and $a\,H$. We show the critical momentum $k_c$, defined in (\ref{kcdef}), for which, the mass and momentum contributions are equal at the time of re-entry, while $k_0$, defined in (\ref{k0def}), is the momentum of the mode for which the mass term became important just today.
\label{Fig:kcdef}}
\end{figure}

We define two special scales which characterize classification of the
different modes. First, we introduce the critical momentum $k_c$ for
which, the horizon crossing occurs at time $t=t_c$, when both the
momentum and the mass term contribute equally to the frequency, 
\begin{equation}
k_c=a_cM_{GW} (t_c) =a_cH_c/\sqrt{2}\,,
\label{kcdef}
\end{equation}
where $a_c\equiv a(t_c)$ and $H_c\equiv H(t_c)$. After the time $t_c$,
the sub-horizon evolution of a mode with $k < k_c$ is completely
determined by the mass term. The second scale we define is 
\begin{equation}
k_0 \equiv a_0\,M_{GW}(t_0)\,,
\label{k0def}
\end{equation}
that characterizes the momentum of the mode for which, the mass term
starts to dominate from today on, $t>t_0$. If $k_c < k_0$ (or
equivalently, by Eq.~(\ref{mgwgrows}), $t_c<t_0$), then the modes which
are affected by the mass term will be within the observable universe
today. In Fig.~\ref{Fig:kcdef}, we show the characteristic momenta $k_c$
and $k_0$ in relation to the evolution.

Throughout this paper we assume that
\begin{equation}
 k_c < k_0, \quad a_c < a_0.
\end{equation}
If $a_c < a_0$ is not satisfied,
the effective mass would always be smaller than the Hubble
expansion rate all the way down to the present time, 
and would never
affect cosmological evolution of gravitational waves.
We also assumed $k_c < k_0$ for simplicity.

\begin{figure}[htbp]
 \centering
\subfigure[Short wavelength: $k_0<k$]{
\includegraphics[clip, width=5cm]{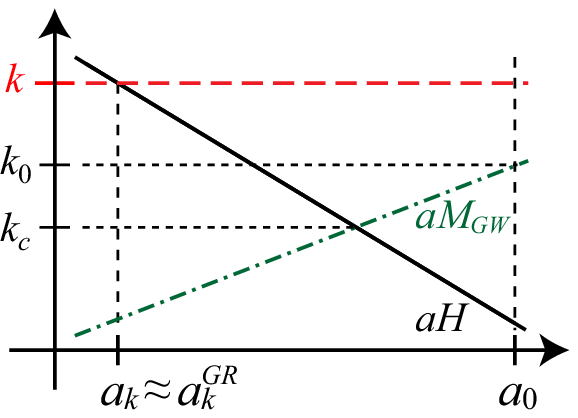}
\label{Fig:H-k-Ma}}
\subfigure[Intermediate wavelength: $k_c < k < k_0$]{
\includegraphics[clip, width=5cm]{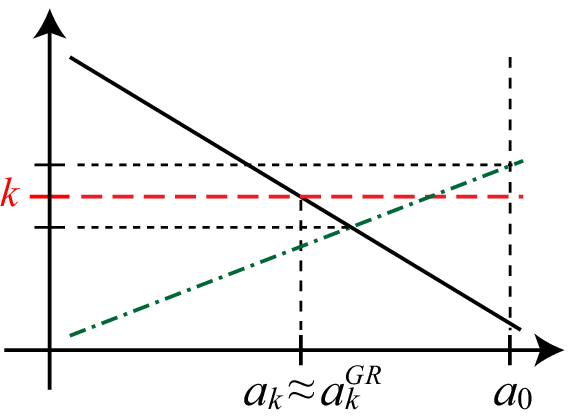}
}
\subfigure[Long wavelength: $k<k_c$]{
\includegraphics[clip, width=5cm]{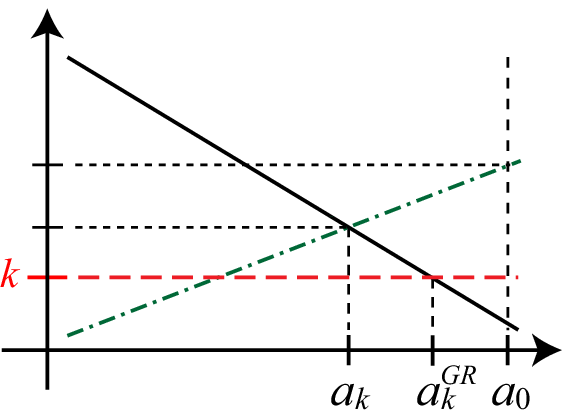}
}
\caption{Examples for each type of momenta based on the classification
 in the main text. For modes with short and intermediate wavelengths ($k_0<k$
 and $k_c<k<k_0$), the horizon entry time is the same as in GR, with  
 $a_k\sim a_k^{GR}$. However, for modes with long wavelength
 ($k<k_c$), the momentum term never becomes dominant and the horizon
 entry in the massive theory occurs earlier than in GR,
 i.e.\ $a_k < a_k^{GR}$. 
}
\label{Fig:H-k-M}
\end{figure}

With these considerations, we classify the modes depending on their momenta:
\begin{itemize}
\item{Short wavelength ($k_0<k$):} modes for which the momentum is
     large enough to dominate the frequency after horizon re-entry. We
     thus have $t_k\simeq t_k^{GR}$, $a_k\simeq a_k^{GR}$ and 
     $H_k\simeq H_k^{GR}$. Since the mass term never becomes important, 
     the evolution is indistinguishable from its counterpart in GR. 
\item{Intermediate wavelength ($k_c<k<k_0$):} modes for which the mass term
     becomes dominant after horizon re-entry, but before today. We still
     have $t_k\simeq t_k^{GR}$, $a_k\simeq a_k^{GR}$ and 
     $H_k\simeq H_k^{GR}$. We expect slight modifications with respect
     to GR signal. 
\item{Long wavelength ($k<k_c$):} modes for which the mass term is
     dominant at horizon re-entry. Since the momentum is negligible
     throughout the evolution after horizon exit, their evolution is the
     same, regardless of their momenta. In other words, $t_k$ for these
     modes is roughly independent of the momentum and all modes of this
     type re-enter the horizon simultaneously. We thus have 
     $t_k\simeq t_c$, $a_k\simeq a_c$ and $H_k\simeq H_c$. The largest
     deviation from general relativity is expected in this category. 
\end{itemize}
This classification is summarized in Fig.~\ref{Fig:H-k-M}.

In the forthcoming calculations, especially in
subsection~\ref{subsec:enhancement}, we will need an explicit 
expression for $a_k^{GR}$. In order to determine this expression, we use 
the Friedmann equation for a composite fluid with nonrelativistic matter
and radiation, that is, 
\begin{equation}
H = \frac{H_{eq}}{\sqrt{2}}\,\left(\frac{a_{eq}}{a}\right)^2\,\sqrt{1+\frac{a}{a_{eq}}}\,,
\end{equation}
where the subscript ``eq'' denotes the time of matter--radiation
equality. (See footnote~\ref{footnote:Lambda-dominance}.) Then, the
horizon crossing time in GR can be found by solving 
\begin{equation}
k^2 = \frac{H_{eq}^2\,a_{eq}^4}{2\,(a_k^{GR})^2}\,
 \left(1+\frac{a_k^{GR}}{a_{eq}}\right)\,.
\end{equation}
Depending on the dominant fluid at $t=t_k^{GR}$, the scale factor
$a_k^{GR}$ will have a different momentum dependence, 
\begin{equation}
 \frac{a_k^{GR}}{a_{eq}} = 
  \left\{
\begin{array}{lll}
\frac{k_{eq}^{GR}}{\sqrt{2}\,k}
 &\quad \mbox{for}\ & k \gg k_{eq} \\
\left(\frac{k_{eq}^{GR}}{\sqrt{2}\,k}\right)^2
 &\quad \mbox{for}\ & k \ll k_{eq} 
      \end{array}
\right.\,,
\label{scalefactor}
\end{equation}
where $k_{eq}^{GR} \equiv a_{eq} H_{eq}$ is the comoving momentum of the
mode which crosses the horizon at the time of equality, in GR.

\subsection{WKB solution}

As already stated in subsection~\ref{subsec:horizon-reentry}, a
super-horizon mode stays constant and, thus, its absolute value is given
by 
\begin{equation}
|\gamma_k| =A(k)\,,\quad {\rm for~}t < t_k\,,
\label{solg1}
\end{equation}
where $A(k)$ is the amplitude of the mode at the time of its generation,
and for slow-roll inflation, is given by 
\begin{equation}
A(k) = \frac{H_*}{M_{Pl}\,k^{3/2}}\,.
\label{primordialAmplitude}
\end{equation}
Here, $H_*$ is the value of the Hubble expansion rate at the time of
horizon exit during inflation.

Once the mode re-enters the horizon at time $t=t_k$, it starts to
oscillate with frequency $\omega$. Assuming that the evolution of the
frequency is adiabatic, i.e.\ $\omega^2\gg H^2$, we can use the WKB
approximation to obtain the solution
\begin{equation}
\gamma_k(t) = \frac{C(k)}{\,\sqrt{\omega(t)a^3(t)}}\,
 \exp\left(-i\,\int \omega\, a\, d\eta\right)\,, \quad {\rm for~}
 t > t_k\, ,
\label{WKBsol}
\end{equation}
where $C(k)$ is a $k$-dependent integration constant. By using the
thin-horizon approximation, with which we suppose the mode re-enters the
horizon sufficiently fast, we can match the super-horizon frozen
solution (\ref{solg1}) and the sub-horizon oscillating solution
(\ref{WKBsol}) at $t=t_k$, resulting in 
\begin{equation}
\frac{\vert \gamma_k (t) \vert}{A(k)} = 
 \sqrt{\frac{\omega_k}{\omega(t)}\frac{a_k^3}{a^3(t)}}
 \,,\quad {\rm for~}t >t_k\,,
\label{gammaMG}
\end{equation}
where $\omega_k\equiv\omega(t_k)=H_k$.

For comparison, the corresponding solution in the massless case (GR) is
obtained by simply replacing $\omega\to k/a$ and $t_k \to t_k^{GR}$ as 
\begin{equation}
\frac{\vert \gamma_k^{ GR} (t) \vert}{A(k)} = 
\frac{a_k^{GR}}{a(t)}
 \,,\quad {\rm for~}t >t_k^{GR}\,.
\label{gammaGR}
\end{equation}

\subsection{Power spectrum}

Observationally speaking, what we are interested in is the amplitude of
sub-horizon gravitational waves today as a function of frequency. 
We thus would like to compare the
amplitude of sub-horizon gravitational waves in massive gravity with
that in GR at the same frequency today. For this purpose we define the 
power spectrum of sub-horizon gravitational waves as 
\begin{equation}
 {\cal P}(\omega_0) \equiv \left.\frac{d}{d\ln \omega_0}
\langle
\gamma_{ij}\gamma^{ij}
\rangle\right|_{t=t_0}
=
\frac{\omega_0^2}{\omega_0^2-M_{GW,0}^2}
\frac{2k^3}{\pi^2}
\left|\gamma_k(t_0)\right|^2,  \quad
k = a_0 \sqrt{\omega_0^2-M_{GW,0}^2}~,
\label{power}
\end{equation}
where 
\begin{equation}
\omega_0 = \omega(t_0), \quad
M_{GW,0} = M_{GW}(t_0),
\end{equation}
and we have used
\begin{equation}
 \frac{d\ln k}{d\ln\omega_0}
  = \frac{\omega_0^2}{\omega_0^2-M_{GW,0}^2}.
\end{equation}
The power spectrum in GR with the same frequency is
\begin{equation}
 {\cal P}_{GR}(\omega_0) = 
\frac{2k'^3}{\pi^2}
\left|\gamma_{k'}^{GR}(t_0)\right|^2,  \quad
k' = a_0 \omega_0\,,
\end{equation}
where $k'$ corresponds to the momentum of the mode in GR with frequency $\omega_0$.

By using the WKB solution given in the previous subsection, we obtain 
\begin{equation}
 {\cal P}(\omega_0) = 
  \left(\frac{k'a_k}{ka_0}\right)^2 
  \frac{\omega_ka_k}{\omega_0a_0}\,
  {\cal P}_{prim}(k), 
\end{equation}
and 
\begin{equation}
 {\cal P}_{GR}(\omega_0) = 
  \left(\frac{a_{k'}^{GR}}{a_0}\right)^2
  {\cal P}_{prim}(k'), 
\end{equation}
where we have defined the primordial power spectrum by
\begin{equation}
 {\cal P}_{prim}(k) \equiv \frac{2k^3}{\pi^2}A^2(k). 
\end{equation}

\subsection{Enhancement of gravitational waves and cutoff}
\label{subsec:enhancement}

We are interested in the ratio of the power spectrum in massive gravity 
to that in GR at the same frequency today $\omega_0$. From the results
in the previous subsection, we obtain
\begin{equation}
 \frac{{\cal P}(\omega_0)}{{\cal P}_{GR}(\omega_0)}
  = \frac{{\cal P}_{prim}(k)}{{\cal P}_{prim}(k')}
  {\cal S}^2(\omega_0),
  \label{eqn:def-A}
\end{equation}
where
\begin{equation}
 {\cal S}(\omega_0) = 
  \frac{k'a_k}{k \,a_{k'}^{GR}}
  \sqrt{\frac{\omega_ka_k}{\omega_0a_0}},  \quad
  k = a_0 \sqrt{\omega_0^2-M_{GW,0}^2}, \quad
  k' = a_0 \omega_0.
\end{equation}
In the rest of this subsection, we estimate the enhancement factor
${\cal S}(\omega_0)$ for the three classes of modes introduced in
subsection~\ref{subsec:classification}.

\subsubsection{Short wavelength modes: $k_0 \ll k$}

For these small-scale gravitational waves, the frequency $\omega$ defined by
(\ref{eqn:dispersion-relation}) is dominated by the momentum throughout
the evolution, i.e.\ $\omega\simeq k/a$. Thus we have 
\begin{equation}
 \omega_k\simeq\frac{k}{a_k}, \quad 
  \omega_0\simeq\frac{k}{a_0}, \quad
  k' \simeq k. 
\end{equation}
Also, as illustrated in Fig.~\ref{Fig:H-k-Ma}, the horizon re-entry
occurs at the same time as in the GR analogue,  
\begin{equation}
 a_k \simeq a_k^{GR} \simeq a_{k'}^{GR}. 
\end{equation}
Thus, this case does not give rise to an enhancement: 
\begin{equation}
 {\cal S}(\omega_0) \simeq 1.
\label{S_short}
\end{equation}

\subsubsection{Intermediate wavelength modes: $k_c \ll k \ll k_0$}

For the intermediate-scale gravitational waves, frequency at the time of
horizon re-entry is dominated by momentum, 
\begin{equation}
 \omega_k \simeq \frac{k}{a_k}\,,
\end{equation}
giving rise to the same re-entry time as in GR
\begin{equation}
 a_k \simeq a_k^{GR}.
\end{equation}
On the other hand, the mass term comes to dominate at later time,
yielding for the frequency of the modes today 
\begin{equation}
 \omega_0 \simeq M_{GW,0} = \frac{k_0}{a_0}, \quad 
  k' \simeq k_0.
\end{equation}
Thus, these modes will undergo an enhancement:
\begin{equation}
 {\cal S}(\omega_0) \simeq 
  \frac{a_k^{GR}}{a_{k_0}^{GR}}
  \sqrt{\frac{k_0}{k}}.
\end{equation}
By using the formula (\ref{scalefactor}), we obtain
\begin{equation}
 {\cal S}(\omega_0) \simeq 
  \left\{
\begin{array}{lll}
 \sqrt{\frac{k_0^3}{k^3}} \simeq
  \left(\frac{\omega_0^2}{M_{GW,0}^2}-1\right)^{-3/4}
 &\quad \mbox{for}\ & k_{eq} \ll k \ll k_0\\
 \sqrt{\frac{k_{eq}^2k_0^3}{2k^5}} \simeq
  \frac{k_{eq}}{\sqrt{2}k_0}
  \left(\frac{\omega_0^2}{M_{GW,0}^2}-1\right)^{-5/4}
 &\quad \mbox{for}\ & k \ll k_{eq} \ll k_0 \\
 \sqrt{\frac{k_0^5}{k^5}} \simeq
  \left(\frac{\omega_0^2}{M_{GW,0}^2}-1\right)^{-5/4}
 &\quad \mbox{for}\ & k \ll k_0\ll k_{eq} 
      \end{array}
\right.\,.
\label{S_int}
\end{equation}

\subsubsection{Long wavelength modes: $k \ll k_c$}

When these modes re-enter the horizon, the mass term is already dominant
in their frequency
\begin{equation}
 \omega_k \simeq M_{GW}(t_k),
\end{equation}
and thus the re-entry time is independent of the comoving momentum. In
other words, all modes with $k \ll k_c$ re-enter the horizon
simultaneously when the Hubble expansion rate and the effective mass of
gravitational waves agree
\begin{equation}
 H(t_k)\simeq M_{GW}(t_k).
\end{equation}
Up to factors of order one, this is essentially the definition of $t_c$ introduced in
subsection~\ref{subsec:classification} and thus $t_k\simeq t_c$. Hence,
we have 
\begin{equation}
 a_k\simeq a_c,\quad \omega_c\simeq M_{GW}(t_c)=\frac{k_c}{a_c}.
\end{equation}
Since we have assumed that $aM_{GW}$ is a growing function of time,
$M_{GW}$ dominates $\omega$ at $t=t_0$ as well: 
\begin{equation}
 \omega_0 \simeq M_{GW,0}=\frac{k_0}{a_0}, \quad k'\simeq k_0. 
\end{equation}
Thus we obtain
\begin{equation}
 {\cal S}(\omega_0) \simeq
  \frac{a_c}{a_{k_0}^{GR}}\frac{\sqrt{k_0k_c}}{k}
  \simeq 
  \frac{a_c}{a_{k_0}^{GR}}
  \sqrt{\frac{k_c}{k_0}}
  \left(\frac{\omega_0^2}{M_{GW,0}^2}-1\right)^{-1/2}. 
  \label{eqn:enhancement}
\end{equation}

\subsubsection{Summary of enhancement factor}
\label{Sec:SummaryOfEnhancementFactor}

\begin{figure}[htbp]
 \centering
\includegraphics[clip, width=5cm]{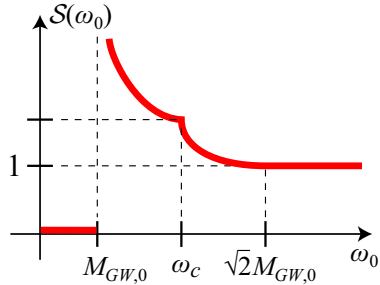}
\caption{Schematic plot of the amplification factor 
${\cal S}(\omega_0)$.}
\label{Fig:DeltaDelta_f}
\end{figure}

As one can see from the schematic plot in Fig.~\ref{Fig:DeltaDelta_f},
qualitative behavior of the amplification factor ${\cal S}(\omega_0)$ is
summarized as follows. 
\begin{itemize}
 \item{No modification in the high frequency range ($k_0<k 
      \,\Leftrightarrow\, \sqrt2 M_{GW,0}<\omega_0$):} 
      ${\cal S}(\omega_0)$ stays almost unity. 
 \item{Modest enhancement in the intermediate frequency range ($k_c<k<k_0 
      \,\Leftrightarrow\,  \omega_c < \omega_0 < \sqrt2 M_{GW,0}$):}
      ${\cal S}(\omega_0)$ is proportional to some positive powers of 
      $(\omega_0^2-M_{GW,0}^2)^{-1}$ 
      as shown in Eq.~(\ref{S_int}),
      and thus increases as 
      $\omega_0$ decreases. 
 \item{Sharp peak just above the cutoff ($0<k<k_c 
      \,\Leftrightarrow\,  M_{GW,0}<\omega_0<\omega_c$):}
      ${\cal S}(\omega_0)$ is proportional to 
      $(\omega_0^2-M_{GW,0}^2)^{-1/2}$, and thus it diverges in the
      limit $\omega_0\to M_{GW,0}$. 
 \item{No signal below the cutoff ($\omega_0<M_{GW,0}$):} 
      ${\cal S}(\omega_0)=0$. 
\end{itemize}
We emphasize that the qualitative feature of the spectrum shown in
Fig.~\ref{Fig:DeltaDelta_f} is quite universal; as long as $a\,M_{GW}$
is an increasing function of time and the two scales $k_c<k_0$ are well
separated, it applies to any fiducial metric, or even other models of
massive gravity. 

Before going into the next subsection, we briefly comment on the case 
where $c_g$ is not constant.%
\footnote{
Note that $c_g$ is bounded from below as $c_g-1 \gtrsim - 10^{-15}$
if we take into account the gravitational Cherenkov radiation~\cite{Moore:2001bv,Kimura:2011qn}.
See also, e.g., Refs.~\cite{Horndeski,Kobayashi:2011nu} for models which may give $c_g\neq 1$.}
We assume that $c_g k/a$ becomes equal to each of $H$ and $M_{GW}$ only once,
similarly to the case of constant $c_g$ we have discussed.
In this case, Eqs.~(\ref{eqn:dispersion-relation}) are altered into
(\ref{kcdef}) and (\ref{k0def})
into
\begin{equation}
\omega^2 = \frac{c_g(t)k^2}{a^2} + M_{GW}^2,
\qquad
c_g(t_c) k_c = a_c M_{GW}(t_c) = a_c H_c,
\qquad
c_g(t_0) k_0 = a_0 M_{GW}(t_0).
\end{equation}
We redefine the horizon crossing time $t_k$ using $\omega$ in this equation.
Using these new definitions, we may repeat the derivations in the previous 
subsections to obtain the formula for ${\cal S}(\omega_0)$.
Its expression is given by
\begin{equation}
{\cal S}(\omega_0) 
=
\frac{k'a_k}{c_g(t_0)k \, a_{k'}^{GR}}
\sqrt{
\frac{\omega_k a_k}{\omega_0 a_0}
}
\simeq
\begin{cases}
\frac{a_k}{a^{GR}_{c_g(t_0)k}}
\sqrt{
\frac{c_g(t_k)}{c_g(t_0)}
} 
& (k_0\ll k)
\\
\frac{a_k}{a^{GR}_{c_g(t_0)k_0}}
\sqrt{
\frac{c_g(t_k)k_0}{c_g(t_0)k}
} 
& (k_c\ll k \ll k_0)
\\
\frac{a_c}{a^{GR}_{c_g(t_0)k}}
\sqrt{
\frac{c_g(t_c)k_0k_c}{c_g(t_0)k^2}
}
& (k \ll k_c)
\end{cases}
~.
\label{S}
\end{equation}
Note that also the primordial spectrum ${\cal P}_{prim}(k)$ 
may be shifted if $c_g(t)$ is shifted from the unity in the inflationary era.

\subsection{Signatures in the spectrum}
\label{subsec:signature}
In the previous subsection, we have shown that the enhancement factor
${\cal S}(\omega_0)$ defined by (\ref{eqn:def-A}) diverges in the limit
$\omega_0\to M_{GW,0}+0$. However, the power spectrum 
${\cal P}(\omega_0)$ itself does not diverge, provided that 
\begin{equation}
\lim_{k\to +0}k^{-2}{\cal P}_{prim}(k)<\infty.
\label{eqn:falloff}
\end{equation}
This is the case, for example, if the primordial gravitational waves are
generated by inflation in the early universe and if the e-folding number
of the inflation is finite. Nonetheless, from the divergence 
of ${\cal S}(\omega_0)$ it is likely that the power spectrum 
${\cal P}(\omega_0)$ has a sharp peak just above the cutoff at
$\omega_0=M_{GW,0}$. Below the cutoff, the power spectrum should vanish.

The peak just above $\omega_0=M_{GW,0}$ may become
  rather high if the primordial spectrum ${\cal P}_{prim}(k)$ of
  gravitational waves continues to a far infrared.  For example, this
  situation can be realized when the duration of inflation is prolonged. 
Concretely, ${\cal P}_{prim}(k)$ is expected to continue
  down to $k\sim a_0H_0e^{-N_{extra}}$, where $N_{extra}$ is the
  number of additional e-folds to the minimum needed for the spatial curvature to stay below the observational bounds. Hence, by the formula
  (\ref{eqn:enhancement}), the enhancement factor at the peak is
  estimated as
\begin{equation}
  {\cal S}_{peak} 
  \sim \frac{a_c}{a_{k_0}^{GR}}
  \frac{\sqrt{k_0k_c}}{a_0H_0}e^{N_{extra}}.
\label{PeakHeight}
\end{equation}

Future observation of gravitational waves may reveal deviation from
the spectrum predicted by GR. If the cutoff and the sharp peak just
above it are found, then this might be a signature of massive gravity. In 
this case, from the frequency at the cutoff one can read off the value
of $M_{GW,0}$. It might also be possible to read off some information
about the fall-off behavior of the primordial spectrum in the far
infrared (\ref{eqn:falloff}), from the height of the peak.

\section{Numerical results}
\label{Sec:example}

In the previous section we have shown that the power spectrum of
stochastic gravitational waves in massive gravity vanishes at
frequencies below a cutoff and a sharp peak is expected just above the
cutoff. The cutoff frequency agrees with the effective mass of
gravitational waves today $M_{GW,0}$. It is therefore interesting if
the effective mass $M_{GW,0}$ is within the frequency range to which
gravitational wave detectors are sensitive. For example, 
evolved Laser Interferometer Space Antenna (eLISA)~\cite{AmaroSeoane:2012km}
is sensitive to the frequency
range $10^{-4}-1$ Hz and aims to be sensitive
for $3\times 10^{-5} - 1$ Hz.
For lower frequencies, pulsar timing
arrays such as the Square Kilometre Array (SKA)~\cite{SKA} and the
Parkes Pulsar Timing Array (PPTA)~\cite{PPTA} have
sensitivities 
down to $10^{-8}$ Hz.  The combined sensitivity range for these
observatories is thus $10^{-8}-1$ Hz.  
Taking into account the bound 
$M_{GW,0} \lesssim 10^{-5}\,\text{Hz}$
from the orbital decays of binary pulsars~\cite{Finn:2001qi},
one might be able to hope future detection of
the cutoff and the peak in the gravitational wave spectrum if
\begin{equation}
10^{-8} \,\text{Hz} \lesssim M_{GW,0}  \lesssim 10^{-5} \,\text{Hz}.
  \label{eqn:MGWrange}
\end{equation}
The spectrum obtained analytically in the previous section, especially
its peak structure, is not sensitive to the history of $M_{GW}(t)$ in the
early universe but is essentially determined by its present value
$M_{GW,0}$. Thus, for the purpose of numerical calculations, we
consider time-independent $M_{GW}=M_{GW,0}$ in the range~(\ref{eqn:MGWrange}).
\footnote{Note that in Sec.~\ref{Sec:basics} we have assumed that
  propagating degrees of freedom in the gravity sector are the two tensor
  modes only. This means that the scalar and vector parts of linear
  perturbations behave exactly like GR and thus there is no constraint
  from modification of Newtonian potential as far as linear
  perturbation is a good approximation.  
See e.g.\ Refs.~\cite{Clifton:2011jh,deRham:2012fw} for the discussions and
    bound on the graviton mass when those degrees of freedom are taken
    into account.}

\begin{figure}[ht]
\includegraphics[width=.47\textwidth]{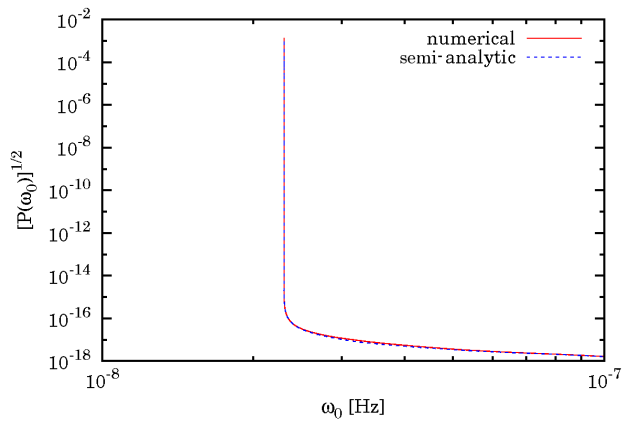}
\caption{ Gravitational power spectrum in the massive gravity with
  $M_{GW,0}=10^{10} H_0\sim 2\times 10^{-7}\,\text{Hz}$ shown with
  respect to $\omega_0$.  The solid (red) curve is the spectrum
  obtained by the numerical calculation, and the dashed (blue) curve
  is the semi-analytical spectrum given by Eq.~(\ref{eqn:def-A}), where
 the GR spectrum is calculated numerically and the enhancement factor 
 ${\cal S}(\omega_0)$ is given analytically by Eqs.~(\ref{S_short}),
 (\ref{S_int}) and (\ref{eqn:enhancement}).  } 
\label{Fig:P-omega}
\end{figure}

In Fig.~\ref{Fig:P-omega}, we compare the numerically obtained
gravitational wave spectrum and the analytic estimation derived in the
previous section.  The former is obtained by numerically solving
Eq.~(\ref{eqtensor}) with the mass term assumed to be 
$M_{GW,0}=10^{10}H_0\sim 2\times 10^{-7}\,\text{Hz}$, 
for which we find $k_0 / H_0 = 10^{10}$ and $k_c/H_0 \sim 10^4$. 
The evolution of the gravitational wave inside the Hubble or mass
horizon is calculated using the WKB solution, Eq.~(\ref{WKBsol}), in
order to save computational time.  The latter is obtained by multiplying
the numerically obtained spectrum in GR by the analytically obtained
factor ${\cal S}(\omega_0)$, given in Eq.~(\ref{S}).  In both cases, we
assume that the evolution of the Hubble expansion rate is determined by
the density parameters of radiation $\Omega_r h^2=4.15\times 10^{-5}$,
matter $\Omega_m h^2=0.1344$, and cosmological constant (either the
genuine cosmological constant or the effective one induced by the
graviton mass term, or their sum) $\Omega_\Lambda=0.734$, with the
normalized Hubble rate $h=0.71$ \cite{Komatsu:2010fb}. The primordial
spectrum is assumed to be scale-invariant and its amplitude is
normalized by ${\cal P}_\text{prim}(k)= 2.43\times 10^{-10}$, where we
assumed the tensor-to-scalar ratio to be $0.1$ and the amplitude of the
scalar primordial spectrum to be $2.43\times 10^{-9}$
\cite{Komatsu:2010fb}. We also suppose that the primordial spectrum have
an IR cutoff at  $k/(a_0H_0)=e^{-N_{extra}}=10^{-1}$.
We can clearly see that the numerical result matches almost exactly the
analytic result derived in Sec.~\ref{Sec:power}, and the peak height is
well described by Eq.~(\ref{PeakHeight}), which gives 
${\cal S}_{peak}\sim 10^{15}$.

\begin{figure}[ht]
\includegraphics[width=.47\textwidth]{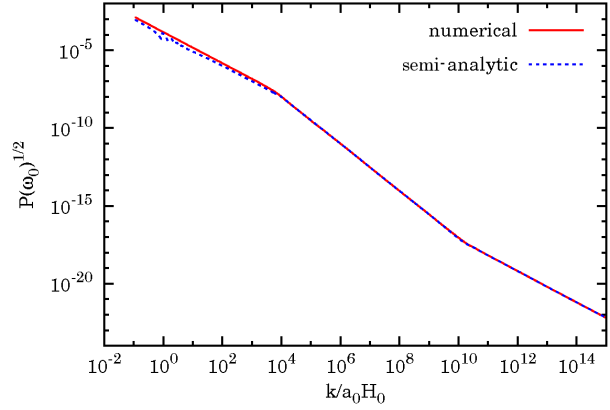}
\caption{ 
 The gravitational power spectra in the massive gravity theory with
 $M_{GW,0}=10^{10} H_0\sim 2\times 10^{-7}\,\text{Hz}$ shown with
 respect to $k/a_0H_0$.  The spectra are obtained by the numerical
 calculation (solid red) and semi-analytic estimate by
 Eqs.~(\ref{eqn:def-A}), (\ref{S_short}), (\ref{S_int}) and
 (\ref{eqn:enhancement}) (dotted blue), respectively. For the latter,
 the GR spectrum is calculated numerically and the enhancement factor
 ${\cal S}(\omega_0)$ is calculated analytically. 
 }
\label{Fig:P-k}
\end{figure}

Since the plot in Fig.~\ref{Fig:P-omega} is shown with respect to
$\omega_0 \sim [ (k/a_0)^2 + M_{GW,0}^2 ]^{1/2}$, most of the peak
structure in the spectrum is compressed to a narrow region at
$\omega_0 \sim M_{GW,0}$ and its detailed structure is too subtle to
see.  To make the fine structure of the peak manifest, we show the
plots with respect to $k$ in Fig.~\ref{Fig:P-k}.  We can see that the
$k$ dependence changes at $k_0$ and $k_c$ as prescribed in
Sec.~\ref{Sec:power}, and that the semi-analytical result well
reproduces the numerical one.

\begin{figure}[ht]
\includegraphics[width=.47\textwidth]{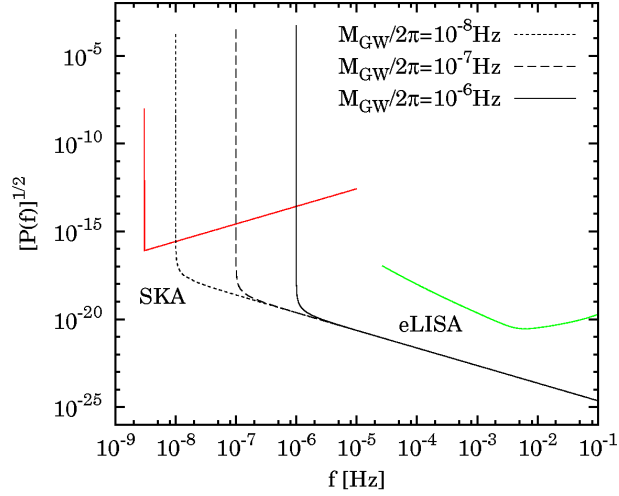}
\caption{ 
Gravitational power spectra for
$M_{GW,0}/2\pi=10^{-8}, 10^{-7}, 10^{-6}\,\text{Hz}$ shown with
sensitivity curves of future experiments.  The curves are plotted
assuming 10-years run of pulsar timing array with SKA~\cite{Sesana:2010mx} 
and 2-years run of eLISA~\cite{AmaroSeoane:2012km}.
    For the primordial
    spectra, we consider the tensor-to-scalar ratio to be $0.1$ and $e^{-N_{extra}}=10^{-1}$. 
}
\label{Fig:sensitivities}
\end{figure}

Before closing this section, we briefly comment on 
  the observability of the peak feature in the spectrum.  As seen from
  Fig.~\ref{Fig:sensitivities}, the peak is observable by the pulsar
  timing 
if the mass is in the range of
  Eq.~(\ref{eqn:MGWrange}).  For example, for the case of
  $M_{GW,0}=10^{-8} \,{\rm Hz}$ and a primordial spectrum with an IR cutoff at
  $k/a_0=1/H_0$, we find that the enhancement factor
  ${\cal S}(\omega_0)$ reaches $\sim 10^{13}$.  Since the enhancement
  is so large, we can expect the peak to be observable even 
  when the tensor-to-scalar ratio takes much smaller value.
The signal might be observable
also for eLISA if its sensitivity could be extended to lower frequencies.

Note that, due to the finiteness of
  the observation time $T_{obs}$, the observed peak in the Fourier
  space is smoothed with a width of $\Delta f\sim 1/T_{obs}$.
The signal would be detectable even after taking this effect into account
if the amplification is sufficiently large.
For example,
following the calculations in Ref.~\cite{Jenet:2005pv} which discussed the pulsar timing observations, 
we find that the total RMS fluctuation of the timing residuals
induced by the stochastic gravitational wave background 
is amplified relative to GR by a factor
$\sim 10^{-4}\times (T_{obs}/H_0)^{-4}(M_{GW,0}/H_0)^{-3}\log(e^{N_{extra}}M_{GW,0}/H_0)$.
For $M_{GW,0}\gtrsim 10^{-8}\,\mathrm{Hz}$,
this factor becomes large enough for the signal to be detectable by the cross correlation analysis
using PPTA argued in Ref.~\cite{Jenet:2005pv}
unless the tensor-to-scalar ratio is extremely small.

\section{Summary and discussions}
\label{Sec:Summary}

In this paper, we discussed how to probe the graviton mass using the
gravitational wave observations. Our results are based on a general
quadratic action given by Eq.~(\ref{I2}) in which the graviton mass can
be time-dependent, and thus they apply universally to any massive
gravity theories in which modification from GR appears primarily in the
tensor modes. The salient feature of the result is the sharp peak in the
gravitational wave spectrum, whose position and height may tell us about
the graviton mass of today and the length of the inflationary period. In
Sec.~\ref{Sec:example}, we confirmed our analytical results match well
with the numerical results which do not assume the thin-horizon
approximation, and also argued the observability of the gravitational
wave signal for various observatories. It would be interesting to apply
the results to explicit examples of nonlinear completion of massive
gravity theories and to derive constraints on such models based on the
gravitational wave observations.

Although the example discussed in Sec.~\ref{Sec:nonlinear} (as well as
in Ref.~\cite{Gumrukcuoglu:2011zh}) was recently shown to suffer from
nonlinear instabilities~\cite{DGM}, it is expected that other working
examples exist. A construction with no extra polarization with respect
to GR was discussed in \cite{Dubovsky:2004sg}, where a constant mass
contribution appears in the dispersion relations of the gravitational
waves. Another potential example is the scenario considered in
\cite{deRham:2012kf}, where the scalar mode is rendered redundant by
introduction of some symmetry. Although the vector graviton degrees are
still present in this construction, the lack of the scalar graviton
would relax the bounds from solar system tests.

Below, we briefly discuss other probes which are sensitive to
$M_{GW}(t)$ at the time other than today.
One interesting probe is the cosmic microwave background (CMB), 
especially its B-mode polarization spectrum
which is sourced solely by gravitational waves.
The signatures in CMB spectrum in the case
of the constant mass $M_{GW}$ was discussed by Ref.~\cite{Dubovsky:2009xk}.
Following the argument to derive the B-mode spectrum 
and generalizing it to the case of time-dependent mass term, 
we find that the main contribution to the spectrum is sensitive 
to the value of the mass at the time of the recombination%
\footnote{We find the contribution of the gravitational wave
to the CMB B-mode spectrum is roughly proportional to 
$\int^{\eta_r} \left(k^2+a^2M_{GW}(\eta)\right)d\eta$, where 
$\eta_r$ is the conformal time at the time of the recombination.
In a generic case, this value is sensitive to $M_{GW}(\eta_r)$ rather
than $M_{GW}(\eta)$ at any other $\eta$.}. 
The most interesting observational feature will be obtained when
the graviton mass at that time is in the range
$(10\text{Mpc})^{-1}< M_{GW} < (10\text{Kpc})^{-1}$.
For such a value of $M_{GW}$, 
the contribution to the B-mode spectrum from the gravitational wave 
increases compared to 
the GR case for the modes of low angular multipole $\ell$, and
a plateau in the spectrum up to 
$\ell\lesssim 10^{-3}\times M_{GW}/H_0 \sim 100$ is expected to emerge. 
There will be no signal for smaller values of $M_{GW}$, while for larger
values, the entire spectrum will be suppressed, since the rapid
oscillation of the gravitational wave, driven by the large $M_{GW}$,
will average out the contribution to the CMB polarization spectrum.


Another interesting probe of the modification in tensor sector of gravity is the 
inspiralling compact binaries; the changes in the propagation speed due to the mass term will give rise to a modification in the evolution of the phase of the gravitational waves emitted by those binaries.
Such an effect and the constraint on the constant graviton mass
are argued in Ref.~\cite{Will:1997bb}.
It may be interesting to generalize this analysis to the case of 
time-dependent graviton mass and discuss its observability for the future-planned 
gravitational wave observatories.

Finally, we give a brief comment on the time evolution of the gravitational wave 
energy density and a possible constraint on it.%
\footnote{
See Ref.~\cite{Dubovsky:2004ud} for the arguments on
possible production of abundant gravitational wave energy density
based on an action similar to our low-energy action of Eq.~(\ref{I2}).
}
The energy spectrum of the gravitational wave defined with respect to wavenumber $k$
is given by
\begin{equation}
\Omega_{GW}(k,t) = \frac{1}{\rho_\text{crit}}\frac{d\rho_{GW}}{d\log k}
= \frac{k^3}{12\pi H^2} \left(
N^{-2}\dot\gamma_k^2 + \omega^2 \gamma_k^2
\right)
\propto
k^3 \omega^2 \gamma_k^2,
\end{equation}
where $\omega$ is given by Eq.~(\ref{eqn:dispersion-relation}) and
the final expression is valid for the WKB solution of Eq.~(\ref{WKBsol}).
The ratio of the energy spectrum of the massive gravity relative to 
GR can be obtained following the derivations similar to Sec.~\ref{Sec:power}.
To discuss the constraint on the gravitational wave energy density 
coming from the Big Bang nucleosynthesis (BBN), 
we need to do the analysis taking $t$ to be the time of BBN.
We assume that the gravitational wave primordial amplitude is given by
Eq.~(\ref{primordialAmplitude}), and consider the case that $M_{GW}(t)$
becomes larger than the Hubble scale before BBN. In such a case, 
the ratio is given by
\begin{equation}
\frac{\Omega_{GW}(k)}{\Omega_{GW}^{GR}(k)}
=
\begin{cases}
1 & (k_0<k)\\
k_0/k & (k_c<k<k_0) \\
k_0 / k_c & (k<k_c)
\end{cases}
~,
\end{equation}
where $k_0$ and $k_c$ are defined in a similar manner with Sec.~\ref{Sec:power}
replacing $t_0$ by $t_\text{BBN}$.
Roughly speaking,
the total energy density of the gravitational wave becomes larger than that in GR by a factor of 
$k_0/k_c = M_{GW}(t_\text{BBN})/\sqrt{H_\text{BBN}M_{GW}(t_c)}$
if $k_c$ is sufficiently large.
Since the total energy density must satisfy $\int d(\log k)\Omega_{GW}(k)<10^{-5}$
at least~\cite{Maggiore:1999vm}, this enhancement of the gravitational wave energy density
may give a tighter constraint on the primordial amplitude of the gravitational wave
or the inflationary models from which it is derived.
Obviously this constraint will be sensitive to $M_{GW}(t_\text{BBN})$.
It may be interesting to pursue this issue 
using
explicit examples
of the massive gravity theories.

\begin{acknowledgments}
\quad
We thank 
D.~Blas,
G.~D'Amico, 
C.~de~Rham, 
S.~Dubovsky,
N.~Kaloper, 
A.~Nishizawa,
A.~Taruya and A.~J.~Tolley for useful discussions. We are also grateful to the organizers and participants of the YITP workshop YITP-T-12-04 ``Nonlinear massive gravity theory and its observational test''
for hospitality and useful comments. The work of A.E.G., C.L.\ and S.M.\ is supported by the World Premier International Research 
Center Initiative (WPI Initiative), MEXT, Japan. S.M. also acknowledges the support by Grant-in-Aid for Scientific Research 
24540256 and 21111006, and by Japan-Russia Research Cooperative Program. N.T.\ is supported in part by the DOE Grant DE-FG03-91ER40674.
S.K.\ is supported by the Grant-in-Aid for Scientific Research No.~23340058 and No.~24740149.
\end{acknowledgments}

\appendix

\section{Example from nonlinear massive gravity}

In the nonlinear theory of massive gravity, the mass term not only
introduces the effective mass to gravitational waves but also drives
self-accelerating universes. In order to attribute the late-time
acceleration of our universe to the self-acceleration from massive 
gravity, the effective cosmological constant $\Lambda_{\pm}$ due 
to the mass term should be set to the observed value:
\begin{equation}
 \Lambda_{\pm} \sim H_0^2. \label{eqn:Lambdatoday}
\end{equation}
Therefore, a relevant question now is whether the nonlinear theory of
massive gravity can simultaneously accommodate the effective mass in the
preferred range (\ref{eqn:MGWrange}) and the preferred value of the
effective cosmological constant (\ref{eqn:Lambdatoday}). These two
quantities are given by the formulas (\ref{mgwdef}) and
(\ref{effectiveLambda}), respectively. (For the sign of the effective
cosmological constant in each branch, see Figure 1 of
Ref.~\cite{Gumrukcuoglu:2011zh}.) Thus, for generic values of the
parameters $\alpha_3$ and $\alpha_4$, the graviton mass $m_g$ in the
action should be 
\begin{equation}
 m_g \sim H_0, 
\end{equation}
and the quantity $r=H/(X_{\pm}H_f)$ at present must be in the range 
\begin{equation}
 10^{20}< r_0 < 10^{28}. \label{eqn:r-range}
\end{equation}
where $r_0\equiv r(t_0)$.

If we suppose that the primordial spectrum ${\cal P}_{prim}(k)$ of
gravitational waves is generated by inflation then in order for the
inflationary mechanism of generation of quantum fluctuations to work, 
the effective mass of gravitational waves must be sufficiently lower
than the Hubble expansion rate during inflation. This is translated to
the condition 
\begin{equation}
 r \ll \frac{H_I^2}{H_0^2} \quad \mbox{during inflation},
  \label{eqn:r-inflation}
\end{equation}
where $H_I$ is the Hubble expansion rate during inflation.

For example, if we demand the Poincar\'e symmetry in the field space as a
global symmetry of the theory then the fiducial metric is restricted to
Minkowski and the FRW background of the physical metric is inevitably an
open universe~\cite{Gumrukcuoglu:2011ew}. In this case we have
\begin{equation}
 r = \frac{Ha}{\sqrt{|K|}}, \label{eqn:r-Minkowski}
\end{equation}
where $K$ ($<0$) is the spatial curvature constant. It is evident that
$r$ increases during inflation and decreases after that (see
footnote~\ref{footnote:Lambda-dominance}). Hence, the condition
(\ref{eqn:r-range}) implies that $r\gg 1$ throughout the evolution after
inflation. In this case $M_{GW}^2\sim r\, m_g^2$ for the '+' branch (or
$M_{GW}^2\sim -r\, m_g^2$ for the '-' branch), and the assumption
(\ref{mgwgrows}) is actually satisfied, provided that $P\leq \rho$ and
$m_g^2>0$ (or $m_g^2<0$, respectively).

Since $r$ decreases after inflation, the condition
(\ref{eqn:r-inflation}) leads to an upper bound on $r_0$. We thus need 
to see if such an upper bound on $r_0$ is consistent with the preferred
range (\ref{eqn:r-range}). For this purpose, supposing that the universe
is dominated by inflaton oscillation (with the effective equation of
state parameter $w=0$) between the end of inflation ($a=a_f$) and the
onset of the radiation dominated epoch ($a=a_R$), the scale factor today
is estimated by the entropy conservation as
\begin{equation}
 \frac{a_0}{a_f} \sim g^{1/12}\frac{\sqrt{M_{Pl}H_I}}{T_0}
  \left(\frac{a_R}{a_f}\right)^{1/4}, \label{eqn:a0}
\end{equation}
where $T_0$ is the photon temperature today and $g$ is the number of
relativistic degrees of freedom that eventually inject entropy to
photons. By using this formula, the bound (\ref{eqn:r-inflation})
at the end of inflation $t=t_f$ is rewritten as
\begin{equation}
 r_0 = r_f\frac{H_0}{H_I}\frac{a_0}{a_f} 
  \ll \frac{H_I}{H_0}\frac{a_0}{a_f}
  \sim g^{1/12}\frac{\sqrt{M_{Pl}H_I}}{T_0}
  \frac{H_I}{H_0}
  \left(\frac{a_R}{a_f}\right)^{1/4}, 
\end{equation}
where $r_f=r(t_f)$. For example, for $g^{1/12}=O(1)$ with the
instantaneous reheating approximation $a_R\simeq a_f$, this leads to 
\begin{equation}
 r_0 \ll 10^{84}\times \left(\frac{H_I}{10^{13}GeV}\right)^{3/2},
\end{equation}
and is consistent with the preferred range (\ref{eqn:r-range}). Thus, in
this simple example with Minkowski fiducial metric, the range
(\ref{eqn:r-range}) is consistent with the inflationary generation
mechanism of primordial gravitational waves.

Note that, by the formula (\ref{eqn:r-Minkowski}), the extreme hugeness
of $r_0$ implied by the preferred range (\ref{eqn:r-range}) is
translated to the extreme flatness of our universe today:
$10^{-56}<|\Omega_{tot}-1|<10^{-40}$. Thus, the number of e-folds during
inflation must be much more than what is needed to solve the flatness
problem. The preferred range (\ref{eqn:r-range}) corresponds to the
number of extra e-folds $\sim \ln r_0$ in the range 
\begin{equation}
 45 < N_{extra} < 65.
\end{equation}
With this large number of extra e-folds, the primordial spectrum 
${\cal P}_{prim}(k)$ of gravitational waves is expected to continue to a
far infrared. Therefore, from the discussion in
subsection~\ref{subsec:signature} we conclude that the peak just above
$\omega_0=M_{GW,0}$ should be rather high. Concretely, 
${\cal P}_{prim}(k)$ is expected to continue down to 
$k\sim a_0H_0e^{-N_{extra}}$. By the formula (\ref{eqn:enhancement}), 
this corresponds to the large enhancement at the peak, 
\begin{equation}
 {\cal S}_{peak} 
  \sim \frac{a_c}{a_{k_0}^{GR}}
  \frac{\sqrt{k_0k_c}}{a_0H_0}e^{N_{extra}}
  \sim \frac{a_c}{a_{k_0}^{GR}}
  \frac{\sqrt{k_0k_c}}{a_0H_0}r_0.
\end{equation}

\section{General fiducial metric}
\label{App:fiducials}

\subsection{FRW-type fiducial metric}
\label{App:FRW}

The background equation of motion for a general fiducial metric is derived in
the Appendix~A.1 of Ref.~\cite{Gumrukcuoglu:2011zh}.
We briefly summarize that derivation for reference.

We consider a FRW-type fiducial metric given by
\begin{equation}
f_{\mu\nu}
=
-\hat n^2(\hat\phi^0) \6_\mu \hat\phi^0 \6_\nu \hat\phi^0
+ \hat \alpha^2(\hat \phi^0) \Omega_{ij}^{(K)}
\6_\mu \hat\phi^i \6_\nu \hat\phi^j,
\label{FRWfiducial2}
\end{equation}
where $\hat n$ and $\hat\alpha$ are general functions of $\hat\phi^0$.
Applying the field redefinition given by
$(\hat\phi^0,\hat\phi^i) =(f(\phi^0),\phi^i)$,
where $f$ is a function to be determined,
and taking the gauge $\phi^\mu = x^\mu$,
the fiducial metric of Eq.~(\ref{FRWfiducial2}) becomes
\begin{equation}
 f_{\mu\nu}dx^\mu dx^\nu =
-\hat n^2 \bigl(f(t)\bigr) \dot f(t)^2 dt^2
+ \hat \alpha^2\bigl(f(t)\bigr) \Omega^{(K)}_{ij} dx^i dx^j,
\end{equation}
where the dot ($\,\dot{}\,$) denotes the derivative with respect to $t$.
For this fiducial metric, the matrix $\K^\mu_{~\nu}$ becomes
\begin{equation}
 \K^\mu_{~\nu} = \text{Diag}\left(\zeta, \xi, \xi, \xi\right),
\qquad
 \zeta = 1-\frac{\hat n \dot f}{N},
\qquad
\xi = 1 - \frac{\hat \alpha}{a}.
\label{zetaxi}
\end{equation}
Then, up to boundary terms the gravity action reduces to
\begin{equation}
 I_g = \int d^3 x \sqrt\Omega \left(
3KNa  -\frac{3a \dot a^2}{N} + m_g^2  L_g
\right),
\label{action_app}
\end{equation}
where
\begin{equation}
L_g = Na^3 \left(
\mathcal L_2 + \alpha_3 \mathcal L_3 + \alpha_4 \mathcal L_4
\right),
\qquad
 \mathcal L_2 = 3\xi(\zeta+\xi),
\qquad
 \mathcal L_3 = \xi^2(3\zeta + \xi),
\qquad
 \mathcal L_4 = \zeta \xi^3.
\end{equation}
The variation of this action with respect to $f$ yields
\begin{align}
0 &=
\frac{\6 L_g}{\6 \zeta}\frac{\6\zeta}{\6 f}
+
\frac{\6 L_g}{\6 \xi}\frac{\6\xi}{\6 f}
- \left(\frac{\6 L_g}{\6 \zeta}\frac{\6 \zeta}{\6 \dot f}\right)^\text{\large
 $\cdot$}%
=
3 Na^2 \left(
aH  - \alpha \hat H_f
\right)
\left[
1+2\xi + \alpha_3(2+\xi)\xi + \alpha_4 \xi^2
\right],
\label{feq}
\end{align}
where
$H$ and $H_f$
are the Hubble parameters of the physical and fiducial metrics
defined by
\begin{equation}
H(t) \equiv \frac{\dot a}{Na},
\qquad
\hat H_f\bigl(f(t)\bigr) \equiv \frac{\hat \alpha'}{\hat n \hat \alpha}
\quad~~
\left(\hat \alpha'\equiv \frac{d\hat \alpha}{d\hat\phi^0}\right).
\end{equation}
Below, we focus on the solution of Eq.~(\ref{feq}) given by%
\footnote{
See Refs.~\cite{Langlois:2012hk, Fasiello:2012rw} for cosmological solutions 
on the other branch.
}
\begin{equation}
 1+ 2\xi + \alpha_3 \xi (2+\xi) + \alpha_4 \xi^2 = 0,
\end{equation}
whose solution is given by
\begin{equation}
 X_\pm = 1-\xi = \frac{\hat \alpha\bigl(f(t)\bigr)}{a(t)},
\qquad
X_\pm \equiv \frac{
1 + 2\alpha_3 + \alpha_4 \pm \sqrt{1+\alpha_3+\alpha_3^2-\alpha_4}
}{\alpha_3+\alpha_4}.
\label{secondsol}
\end{equation}
Varying the action with respect to $N$ and $a$
and using the solution of Eq.~(\ref{secondsol}),
we obtain
Eq.~(\ref{Friedmann}) as the background equations.
The second equation in Eq.~(\ref{Friedmann}) is consistent with the first one if
the matter fluid obeys usual conservation law
$\dot \rho + 3(\dot a/a)(\rho+p)=0$.
The evolution of $a$ is determined by Eq.~(\ref{Friedmann}) for given $N$,
and it determines
$f(t)$ via Eq.~(\ref{secondsol}). Note that $\hat n (f(t))$,
$\hat \alpha (f(t))$ and $\hat H_f(f(t))$
correspond to $n(t)$, $\alpha(t)$ and $H_f(t)$
of Sec.~\ref{Sec:basics}, respectively.
(Almost equivalently, we may set $f(t)=t$ and regard $\hat \alpha(t)$ and $\hat n(t)$
are given functions. In this case, Eq.~(\ref{secondsol}) and the Friedmann
equation in Eq.~(\ref{Friedmann})
are regarded as equations to fix $a(t)$ and $N(t)$, respectively.)

One example of the fiducial metric of this type
is the Minkowski metric
in the open chart ($K<0$) given by~\cite{Gumrukcuoglu:2011ew}
\begin{equation}
f_{\mu\nu} =
-\6_\mu\varphi^0  \6_\nu\varphi^0
+ \delta_{ij} \6_\mu \varphi^i \6_\nu \varphi^j
=
-{f'}^2(\phi^0) \6_\mu \phi^0 \6_\nu \phi^0
+ |K|f^2(\phi^0) \Omega_{ij}^{(K)} dx^i dx^j,
\end{equation}
where we used the field redefinition given by
\begin{equation}
 \varphi^0 = f(\phi^0)\sqrt{1-K\delta_{ij}\phi^i\phi^j},
\qquad
\varphi^i = \sqrt{-K}f(\phi^0) \phi^i.
\end{equation}
Taking the gauge $\phi^i=x^i$, we find Eq.~(\ref{secondsol}) becomes
$X_\pm = \sqrt{|K|}f(t)/a(t)$, which determines $f(t)$ in terms of $a(t)$.
The Friedmann equation, Eq.~(\ref{Friedmann}), holds as it is.
In this case we have $n(\phi^0) = |f'(\phi^0)|$ and
$\alpha(\phi^0)= \sqrt{|K|}f(\phi^0)$,
and thus $H_f = \text{sign}(f')/f = \text{sign}(f')\sqrt{|K|}/X_\pm a$.

\subsection{(Anti-)de Sitter fiducial metric}
\label{App:dS}

In this section, we comment that the fiducial metric of Eq.~(\ref{FRWfiducial})
for any $K$ can be obtained from a single de Sitter metric,
and the for $K>0$ can be obtained from the anti-de Sitter metric.

We begin with the de Sitter fiducial metric with flat spatial section
given by
\begin{equation}
 f_{\mu\nu} =
- \6_\mu \varphi^0 \6_\nu \varphi^0
+ \exp\bigl(2\bar H_f {\varphi^0} \bigr)
\delta_{ij} \6_\mu \varphi^i \6_\nu \varphi^j.
\label{dSfiducial}
\end{equation}
By the redefinition $\varphi^0=f(\phi^0)$ and $\varphi^i=\phi^i$,
this fiducial metric takes
the form of Eq.~(\ref{FRWfiducial}) with $K=0$,
$H_f=\bar H_f$, $n=\bigl|\dot f\bigr|$ and
$\alpha = \exp\bigl(2\bar H_f f(\phi^0)\bigr)$.

Assuming $K>0$, the field redefinition given by
\begin{equation}
\varphi^0 = \bar H_f^{-1} \log\left[
\sinh f(\phi^0)
+ \cosh f(\phi^0) \sqrt{1-K\delta_{ij}\phi^i\phi^j}
\right],
\qquad
\varphi^i =
\frac{
\bar H_f^{-1} \sqrt{K}\phi_i
}{
\tanh f(\phi^0) + \sqrt{1-K\delta_{ij}\phi^i\phi^j}
}
\end{equation}
makes the fiducial metric of Eq.~(\ref{dSfiducial}) into
\begin{equation}
f_{\mu\nu} =
\bar H_f^{-2}\left(
- {f'}^2(\phi^0)
\6_\mu \phi^0 \6_\nu \phi^0
+ K \cosh^2 \! f(\phi^0)
\,
\Omega^{(K)}_{ij} \6_\mu \phi^i \6_\nu \phi^j
\right),
\label{dSfiducial_closed}
\end{equation}
which is of the form of Eq.~(\ref{FRWfiducial}) with $K>0$
and $H_f=\bar H_f \tanh f(\phi^0)$.
The counterpart of Eq.~(\ref{secondsol}) for Eq.~(\ref{dSfiducial_closed})
is given by
\begin{equation}
X_\pm = \frac{\sqrt{K}}{\bar H_f}
\frac{\cosh \bigl(f(t)\bigr)}{a(t)}
,
\label{constraint_dSclosed}
\end{equation}
and the Friedmann equation, Eq.~(\ref{Friedmann}), holds as it is.

Similarly, for $K<0$, the redefinition given by
\begin{equation}
\varphi^0 = \bar H_f \log\left[
\cosh f(\phi^0)
+ \sinh f(\phi^0) \sqrt{1-K \delta_{ij}\phi^i\phi^j}
\right],
\qquad
\varphi^i=
\frac{\bar H_f^{-1}\sqrt{|K|}\phi^i}{
\coth f(\phi^0) + \sqrt{1-K\delta_{ij}\phi^i\phi^j}
}
\end{equation}
makes Eq.~(\ref{dSfiducial}) into
\begin{equation}
f_{\mu\nu} =
\bar H_f^{-2}\left[
-
{f'}^2 (\phi^0)
\6_\mu \phi^0 \6_\nu \phi^0
+ |K|\sinh^2 \! f(\phi^0) \,
\Omega^{(K)}_{ij} \6_\mu \phi^i \6_\nu \phi^j
\right],
\end{equation}
which is of the form of Eq.~(\ref{FRWfiducial}) with $K<0$
and $H_f = \bar H_f \coth f(\phi^0)$.
The counterpart of Eq.~(\ref{secondsol}) is given by
\begin{equation}
X_\pm  = \frac{\sqrt{|K|}}{\bar H_f}
\frac{\sinh \bigl(f(t)\bigr)}{a(t)}.
\label{constraint_dSopen}
\end{equation}

Next, we consider the anti-de Sitter (AdS) fiducial metric given by
\begin{equation}
 f_{\mu\nu} = - \left(
1+k^2 \delta_{ij} \varphi^i \varphi^j
\right) \6_\mu \varphi^0 \6_\nu \varphi^0
+
\Omega^{(-k^2)}_{ij} \6_\mu \varphi^i \6_\nu \varphi^j,
\end{equation}
which is the AdS metric with curvature $k$ in the global chart.
Assuming $K<0$, the field redefinition given by
\begin{equation}
 \varphi^0 = k \arcsin\left(
\frac{\sin f(\phi^0)}{\sqrt{1-K\cos^2\! f(\phi^0)\, \delta_{ij}\phi^i \phi^j }}
\right),
\qquad
\varphi^i = \frac{\sqrt{|K|}}{k} \cos\! f(\phi^0)\, \phi^i
\end{equation}
make the fiducial metric into
\begin{equation}
 f_{\mu\nu} =
k^{-2}\left(
- {f'}^2 (\phi^0) \6_\mu \phi^0 \6_\nu \phi^0
+ \left|K\right| \cos^2\! f(\phi^0) \, \Omega^{(K)}_{ij} \6_\mu \phi^i \6_\nu \phi^j
\right),
\end{equation}
for which $H_f = k \tan f(\phi^0)$ and the counterpart of Eq.~(\ref{secondsol})
is
\begin{equation}
 X_\pm = \frac{\sqrt{\left|K\right|}}{k} \frac{\cos f(t)}{a(t)}.
\end{equation}
We summarize these results in Table~\ref{results}.
Note the the constraint equations, which are the counterparts of Eq.~(\ref{secondsol}),
give restrictions on the trajectory of $a(t)$.
For example, the constraint equation in the $K>0$ case allows only the 
trajectory of $a(t)$ which starts from the infinity, decreases to the minimum and 
grows to the infinity again.
Even though the $a(t)$ which starts from zero, increases to the maximum and 
comes back to zero is also allowed by the Friedmann equation, it is not 
consistent with the constraint equation and is not a cosmological solution in 
this model.

\begin{table}[htbp]
 \caption{Results summary.}
\centering
\begin{tabular}{|c|c|c|c|c|l|%
@{\protect\rule[-.55\normalbaselineskip]{0pt}{1.7\normalbaselineskip}}
}
\hline
& &$\varphi^0$ & $\varphi^i$ & $H_f$ &
\multicolumn{1}{|c|}{Constraint eq.}  \\ \hline
&$K>0$
& $\bar H_f^{-1}\log\left[\sinh f(\phi^0) + \cosh f(\phi^0)
\sqrt{1-K\delta_{ij} \phi^i \phi^j}\right]$
& $\frac{\bar H_f^{-1}\sqrt{K} \phi^i}{\tanh f(\phi^0) +
\sqrt{1-K\delta_{ij} \phi^i \phi^j}
}$
& $\bar H_f \tanh\bigl(f(\phi^0)\bigr)$
& $X_\pm = \frac{\sqrt{K}}{\bar H_f}\frac{\cosh f(t)}{a(t)}$
\\ \cline{2-6}
dS
&$K=0$
& $f(\phi^0)$
& $\phi^i$
& $\bar H_f$
& $X_\pm = \frac{\exp(\bar H_f f(t))}{a(t)}$
\\ \cline{2-6}
&$K<0$
& $\bar H_f^{-1}\log\left[\cosh f(\phi^0) + \sinh f(\phi^0)\sqrt{1-K
\delta_{ij} \phi^i \phi^j}\right]$
& $\frac{\bar H_f^{-1}\sqrt{|K|}\phi^i}{\coth f(\phi^0)
+ \sqrt{1-K\delta_{ij}\phi^i \phi^j}}$
& $\bar H_f \coth\bigl(f(\phi^0)\bigr)$
& $X_\pm = \frac{\sqrt{\left|K\right|}}{\bar H_f} \frac{\sinh f(t)}{a(t)}$
\\
\hline
\hline
AdS
&$K<0$
& $k\arcsin
  \left(
   \frac{\sin f(\phi^0)}{\sqrt{1-K\cos^2 f(\phi^0)\delta_{ij}\phi^i\phi^j}}
  \right)$
& $\frac{\sqrt{|K|}}{k}\cos f(\phi^0)\phi^i$
& $k \tan f(\phi^0)$
& $X_\pm = \frac{\sqrt{\left|K\right|}}{k} \frac{\cos f(t)}{a(t)}$
\\
\hline
\end{tabular}
\label{results}
\end{table}

Before closing this appendix, we briefly mention some examples of $M_{GW}$.
For instance, in the case of the $+$ blanch solution with the Minkowski
fiducial metric, 
\begin{equation}
\left.M^2_{GW}\right\vert_{\rm Minkowski} 
 \propto \left(\dot{a} + {\rm constant}\right)\,.
\end{equation}
As a result, $a^2 |M^2_{GW}|$ always grows with time, provided that the
dominant fluid's equation of state is $w <1$. Another example is 
the de Sitter fiducial metric, for which we have 
\begin{equation}
\left.M^2_{GW}\right\vert_{\rm de Sitter} 
 \propto \left(\frac{\dot{a}}{a} + {\rm constant}\right)\,.
\end{equation}
In this case, the contribution from the first term in $a^2 M^2_{GW}$
will grow if $w <1/3$. Although for radiation dominated universe, this
contribution is constant, the second term will still grow as $a^2$. In
the light of these examples and the late-time stability of the system,
we adopt the assumption (\ref{mgwgrows}) throughout this paper.

\section{Analytic solutions of $\gamma_k$}
\label{App:analytic}

We show the analytic solution for Eq.~(\ref{eqtensor}) for power-law type
$M_{GW}^2 = M_\text{eq}^2 (a/a_\text{eq})^p$,
which covers the simple examples such as
the constant $M_{GW}$ 
and 
also the examples mentioned above.
The aim of this section is to confirm the validity of
the thin-horizon approximation used in Sec.~\ref{Sec:power}.

We define the conformal time $\eta$
from the scale factor
$a = a_\text{eq} (t/t_\text{eq})^{1/2}$ (RD) and
$a = a_\text{eq} (t/t_\text{eq})^{3/2}$ (MD) as
\begin{equation}
 \eta \equiv \alpha \int_0^t \frac{dt}{a(t)} =
\eta_\text{eq} \times
\begin{cases}
 \frac{a}{a_\text{eq}}
& (\text{RD})
\\
\frac23 \left(\frac{a}{a_\text{eq}}\right)^{1/2} - \frac12
& (\text{MD})
\end{cases}
\qquad \Leftrightarrow \qquad
\frac{a}{a_\text{eq}} =
\begin{cases}
 \frac{\eta}{\eta_\text{eq}}& (\text{RD}) \\
\left(\frac23\left(\frac{\eta}{\eta_\text{eq}}+\frac12\right)\right)^2
& (\text{MD})
\end{cases}
~,
\label{eta}
\end{equation}
where $\alpha$ is a constant and
$\eta_\text{eq} \equiv 2\alpha t_\text{eq}/a_\text{eq}$.

We consider the case that $M_{GW}$ becomes dominant in the MD era
first. Eq.~(\ref{eqtensor}) in this case becomes
\begin{equation}
\bar\gamma_k'' + \left(k^2 + a^2 M_\text{eq}^2
\left(\frac{a}{a_\text{eq}}\right)^p
  - \frac{a''}{a}\right)\bar\gamma_k = 0
\qquad
\Leftrightarrow
\qquad
\frac{d^2 \bar\gamma_k}{d\hat\eta^2} + \left(
\hat k^2 + \left(\frac23\right)^{2(p+2)}
\hat M_\text{eq}^2 \hat \eta^{2(p+2)}
- \frac{2}{\hat \eta^2}
\right)\bar\gamma_k = 0,
\label{eqtensor_MD}
\end{equation}
where we introduced
$\hat k \equiv k \eta_\text{eq}$,
$\hat M_\text{eq} \equiv a_\text{eq}\eta_\text{eq}M_\text{eq}$,
$\hat a \equiv a / a_\text{eq}$ and
$\hat \eta \equiv \frac{\eta}{\eta_\text{eq}} + \frac12
= 2\hat a^{1/2}/3$.
Below, we use an approximation depending on which of
$k$ and $a M_{GW}$ is dominant at the moment of the horizon crossing.
\begin{enumerate}
 \item $ k \gg aM_{GW}$ at the horizon crossing:

In this case, Eq.~(\ref{eqtensor_MD}) is approximated near the moment of
the horizon crossing as
\begin{equation}
 \frac{d^2 \bar\gamma_k}{d\hat \eta^2} + \left(
\hat k^2 - \frac{2}{\hat\eta^2}
\right)\bar\gamma_k = 0,
\end{equation}
which is same as that in the pure GR.
This equation has an exact solution
which becomes regular for $\hat k \hat \eta \to 0$
given by
\begin{equation}
 \bar\gamma_k = C\times \frac{3}{\hat k^2}\left(
-\cos\bigl(\hat k\hat \eta\bigr)
+ \frac{\sin\bigl(\hat k\hat \eta\bigr)}{\hat k \hat \eta}
\right) \bar\gamma_k
\xrightarrow{\hat k\hat \eta\ll 1} C\hat\eta^2.
\label{gamma_MD_GR}
\end{equation}
The primordial amplitude is given by
$|\gamma_k| = k^{-3/2}(H_*/M_{Pl})$ and
\begin{equation}
 \left|\bar\gamma_k\right|
=  \frac{H_* a}{k^{3/2}M_{Pl}}
=  \frac{4 H_* a_\text{eq}\eta_\text{eq}^{3/2}}
{9M_{Pl}\hat k^{3/2}} \hat \eta^2
\equiv \bar \gamma_* \hat \eta^2
,
\label{primordialbargamma}
\end{equation}
and then we may identify it with
the right-hand side of Eq.~(\ref{gamma_MD_GR})
in the limit of $\hat k\hat\eta \ll 1$.
The solution in the WKB regime,
on the other hand, is given by
\begin{equation}
 \bar\gamma_k = \frac{C'}{\sqrt{\hat{\tilde k}}}
\cos\left(\int \hat {\tilde k} d\hat \eta\right),
\end{equation}
where $\hat{\tilde k}^2\equiv (k^2+a^2M_{GW}^2)\eta_\text{eq}^2$.
Comparing it with Eq.~(\ref{gamma_MD_GR}) in the limit of
$\hat k\hat \eta\gg 1$, we may fix the coefficient $C'$
to have
\begin{equation}
 \bar\gamma_k =
\frac{3\bar\gamma_*}{\hat k^{3/2}}
\cos\left(
\int \hat{\tilde k} d\hat \eta
\right).
\label{gamma_MD_GR2}
\end{equation}

Let us compare Eq.~(\ref{gamma_MD_GR2}) with $\gamma_k$ obtained using
the thin-horizon approximation. The condition to determine $a$ at the horizon
crossing is given by $k = a \hat H = a'/a = 2/\eta$.
Matching the WKB solution with the primordial solution at $\eta$ defined by this
       equation, we find the solution is given by
\begin{equation}
\bar  \gamma_k = 
\frac{4\bar\gamma_*}{\hat k^{3/2}}
\cos\left(\int \hat{\tilde k} d\hat \eta\right),
\end{equation}
which coincide with Eq.~(\ref{gamma_MD_GR2}) up to a numerical factor.

\item $a M_{GW}\gg k$ at the horizon crossing:

Eq.~(\ref{eqtensor_MD}) in this case is approximated as
\begin{equation}
\bar\gamma''_k
+ \left(
a^2 M_{GW}^2 - \frac{a''} {a}
\right)\bar\gamma = 0
\qquad\Leftrightarrow\qquad
\frac{d \bar\gamma}{d\hat \eta^2} + \left(
\left(\frac23\right)^{2(p+2)}
\hat M_\text{eq}^2
 \hat \eta^{2(p+2)}
- \frac{2}{\hat\eta^2}
\right)\bar\gamma_k = 0,
\end{equation}
which have an exact solution that becomes regular for $\hat k \hat \eta\to 0$
given by
\begin{equation}
 \bar\gamma_k =
C\times
\frac{
\left(\frac32\right)^{\frac32-\frac3q}
q^{\frac3q} \Gamma\left(1+\frac3q\right)
}{  \hat M_\text{eq}^{\frac3q} }
~
\hat\eta^{1/2}
 J_{\frac3q}\left(
\frac{3\left(\frac23\right)^{\frac{q}{2}} \hat M_\text{eq}}{q}
\hat \eta^{\frac{q}{2}}
\right)
\xrightarrow{\hat \eta\to 0}
C\hat \eta^2,
\end{equation}
where $J$ is the Bessel function and we have defined $q\equiv 2(p+3)$.
Fixing $C$ using Eq.~(\ref{primordialbargamma}), we find
$\bar\gamma_k$ to be
\begin{multline}
 \bar\gamma_k =
\bar\gamma_*
\times
\frac{
\left(\frac32\right)^{\frac32-\frac3q}
q^{\frac3q} \Gamma\left(1+\frac3q\right)
}{  \hat M_\text{eq}^{\frac3q} }
~
\hat\eta^{1/2}
 J_{\frac3q}\left(
\frac{3\left(\frac23\right)^{\frac{q}{2}} \hat M_\text{eq}}{q}
\hat \eta^{\frac{q}{2}}
\right)
\\
\xrightarrow{\hat \eta\to \infty}
\frac{
q^{\frac3q+\frac12}
\Gamma\left(1+\frac3q\right)
}{\sqrt\pi }
\times
\frac{
\bar\gamma_* \left(\frac32\right)^{\frac32-\frac3q} \hat  M_\text{eq}^{-\frac3q}
}{
\sqrt{
\left(\frac23\right)^{\frac{q}{2}-1}
\hat M_\text{eq} \hat \eta^{\frac{q}{2}-1}
}
}
\cos\left(
\frac{3\left(\frac23\right)^{\frac{q}{2}} \hat M_\text{eq}}{q}
\hat \eta^{\frac{q}{2}}
\right)
,
\label{Exact2WKB}
\end{multline}
where we neglected the phase of the oscillation.
This expression is valid as long as $M_{GW}$ dominates over $k/a$.

Below, we compare it with the expression obtained from the thin-horizon 
approximation. The condition to determine $\eta$ 
at the horizon crossing is given by
\begin{equation}
\left(\frac23\right)^{2(p+2)}\hat M_\text{eq}^2 \hat \eta^{2(p+2)}
= \frac{2}{\hat\eta^2}
\qquad
\therefore
\quad
\eta = 2^{\frac{1}{2(p+3)}}
\left(\frac32\right)^{\frac{p+2}{p+3}}\hat M_\text{eq}^{-\frac{1}{p+3}}.
\end{equation}
Matching the WKB solution with the primordial solution at this $\eta$,
we find 
\begin{equation}
\bar \gamma_k = 
2^{\frac14 + \frac{3}{2q}} 
\times
\frac{
\bar \gamma_* \left(\frac32 \right)^{\frac32 - \frac3q }\hat M_\text{eq}^{-\frac3q}
}{\sqrt{\hat {\tilde k}}}\cos\left(\int\hat{\tilde k} d\hat \eta\right).
\end{equation}
This expression coincides with Eq.~(\ref{Exact2WKB}) 
up to an $\mathcal{O}(1)$ factor
for generic $q>0$,
and thus the thin-horizon approximation is fairly good even in this case.

\end{enumerate}

Finally, we briefly comment on the case that $M_{GW}$ becomes dominant
in the RD era. In this case, Eq.~(\ref{eqtensor}) becomes
$ \bar\gamma'' + \left(k^2 + a^2 M_{GW}^2  \right)\bar\gamma=0$.
As long as $k^2+a^2 M_{GW}^2$ is positive
and its time variation is sufficiently slow,
the solution will be well approximated by the WKB solution.
The amplitude of this solution is determined by matching it to the primordial
solution.
By construction, this solution will be well approximated by the solution given by
the thin-horizon approximation in general.

\end{document}